\pdfoutput=1

\documentclass[11pt]{article}

\usepackage[final]{acl}

\usepackage{times}
\usepackage{latexsym}

\usepackage[T1]{fontenc}

\usepackage[utf8]{inputenc}

\usepackage{microtype}

\usepackage{inconsolata}

\usepackage{graphicx}

\title{SuPreME: A Supervised Pre-training Framework for Multimodal ECG Representation Learning}

\author{
 \textbf{Mingsheng Cai\textsuperscript{1,2}},
 \textbf{Jiuming Jiang\textsuperscript{1,2}},
 \textbf{Wenhao Huang\textsuperscript{3}},
 \textbf{Che Liu\textsuperscript{2\thanks{Correspondence: \href{mailto:che.liu21@imperial.ac.uk}{che.liu21@imperial.ac.uk}}}},
 \textbf{Rossella Arcucci\textsuperscript{2}}
\\
 \textsuperscript{1}The University of Edinburgh,
 \textsuperscript{2}Imperial College London, \\
 \textsuperscript{3}Shenzhen Yinwang Intelligent Technology Co., Ltd
\\
  \texttt{\{mingsheng.cai, jiuming.jiang\}@ed.ac.uk}\textsuperscript{1}, \\
  \texttt{\{mingsheng.cai23, jiuming.jiang23, che.liu21, r.arcucci\}@imperial.ac.uk}\textsuperscript{2}, \\
  \texttt{huangwenhao@yinwang.com}\textsuperscript{3} \\
}

\begin{document}

\maketitle

\begin{abstract}

Cardiovascular diseases are a leading cause of death and disability worldwide. Electrocardiogram (ECG) is critical for diagnosing and monitoring cardiac health, but obtaining large-scale annotated ECG datasets is labor-intensive and time-consuming. Recent ECG Self-Supervised Learning (eSSL) methods mitigate this by learning features without extensive labels but fail to capture fine-grained clinical semantics and require extensive task-specific fine-tuning.
%
To address these challenges, we propose \textbf{SuPreME}, a \textbf{Su}pervised \textbf{Pre}-training framework for \textbf{M}ultimodal \textbf{E}CG representation learning. SuPreME is pre-trained using structured diagnostic labels derived from ECG report entities through a one-time offline extraction with Large Language Models (LLMs), which help denoise, standardize cardiac concepts, and improve clinical representation learning. By fusing ECG signals with textual cardiac queries instead of fixed labels, SuPreME enables zero-shot classification of unseen conditions without further fine-tuning.
We evaluate SuPreME on six downstream datasets covering 106 cardiac conditions, achieving superior zero-shot AUC performance of 77.20\%, surpassing state-of-the-art eSSLs by 4.98\%\footnote{All code and data are available at \url{https://github.com/mingscai/SuPreME}.}. Results demonstrate SuPreME's effectiveness in leveraging structured, clinically relevant knowledge for high-quality ECG representations.

\end{abstract}

\section{Introduction}

Supervised learning methods have proven effective in classifying cardiac conditions using Electrocardiogram (ECG), a widely utilized clinical tool for monitoring the heart's electrical activity \citep{SPN,SPNv2}. However, these methods typically rely on large-scale, high-quality annotated datasets, which are costly to create and difficult to scale.

To reduce dependence on annotations, recent advancements in ECG self-supervised learning (eSSL) have enabled the extraction of representative features from large-scale unannotated ECGs using contrastive or generative tasks \citep{eldele2021time, kiyasseh2021clocs, na2024guiding}. Despite their promise, these methods often rely on strong signal-level augmentations that may distort the semantic integrity of the signal and require complex pretext task designs \citep{kiyasseh2021clocs}.
Multimodal learning approaches \citep{liu2024zero, li2024frozen} have also been proposed to learn ECG representations by leveraging free-text ECG reports. However, these methods face challenges due to noise in textual data and the complexities of language grammar, which can hinder learning efficiency \citep{wu2023medklip}.

\begin{figure}[tp!]
    \includegraphics[width=\columnwidth]{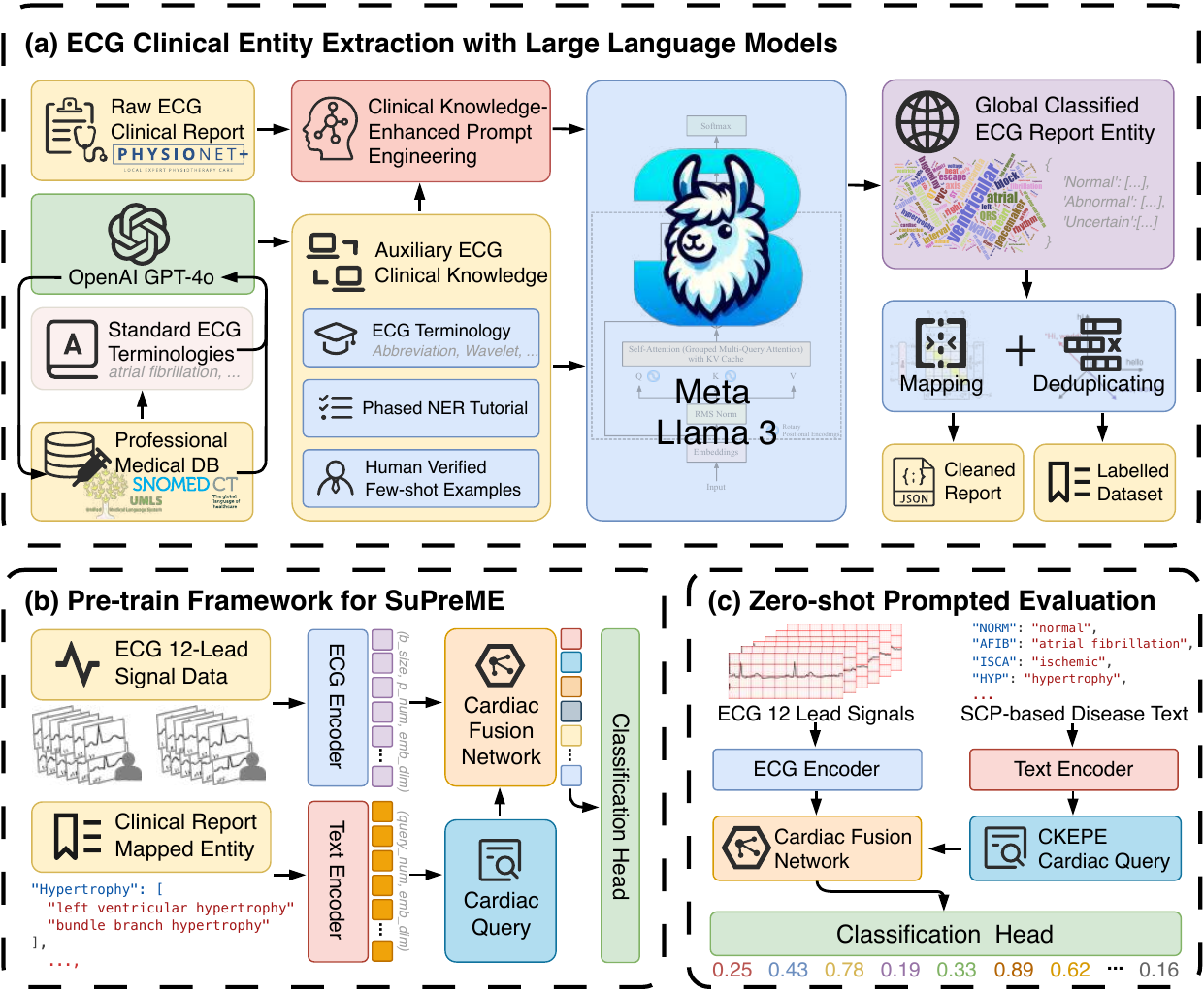}
    \vspace{-20pt}
    \caption{Overview including (a) ECG clinical entity extraction with LLMs, (b) supervised pre-training with SuPreME, and (c) zero-shot prompted evaluation.}
    \label{fig:framework}
    \vspace{-20pt}
\end{figure}

To address these limitations and develop a scalable, simple, and effective ECG pre-training framework, we propose \textbf{SuPreME}, a \textbf{Su}pervised \textbf{Pre}-training framework for \textbf{M}ultimodal \textbf{E}CG representation learning. Our contributions are threefold:
\textbf{(a)} We introduce an automated pipeline that extracts high-quality clinical entities from raw ECG reports using an instruction-tuned LLM enriched with domain-specific knowledge (Figure~\ref{fig:framework}(a)). Extracted entities are deduplicated and mapped to dataset-specific standardized diagnostic terms using clinician-validated resources (e.g., SNOMED CT, UMLS, SCP-ECG\footnote{SNOMED CT and UMLS are standardized clinical terminology databases; SCP-ECG refers to the Standard Communication Protocol for Computer-Assisted Electrocardiography.}), forming a dataset-specific global cardiac query list without manual annotation. This process enables scalable, consistent labeling and captures richer semantics than coarse-grained or free-text alternatives.
\textbf{(b)} Leveraging these standardized cardiac queries, we propose SuPreME (Figure~\ref{fig:framework}(b-c)), a multimodal framework that directly fuses ECG signals with cardiac queries through a lightweight Cardiac Fusion Network (CFN). Unlike prior methods (e.g., MERL) that rely on raw free-text inputs or handcrafted pretext tasks, SuPreME requires no signal-level augmentation or contrastive loss design, offering an efficient and interpretable multi-label supervision strategy grounded in standardized cardiac queries.
\textbf{(c)} We pre-train SuPreME on 771,500 ECG signals paired with 295 global standardized cardiac queries from MIMIC-IV-ECG~\citep{gow2023mimic} (Appendix \ref{app:data-model-pretrain}). On six downstream datasets (e.g., PTB-XL, CPSC-2018, Chapman-Shaoxing-Ningbo; Appendix \ref{app:data-model-downstream}), it achieves a new state-of-the-art zero-shot AUC of 77.20\%, significantly outperforming existing eSSL and multimodal baselines, including those fine-tuned with 10–100\% labeled data. SuPreME also shows strong data efficiency and generalization, with zero-shot performance under only 20\% pre-training data surpassing fully fine-tuned eSSLs.

\begin{figure}[htbp]
    \includegraphics[width=\columnwidth]{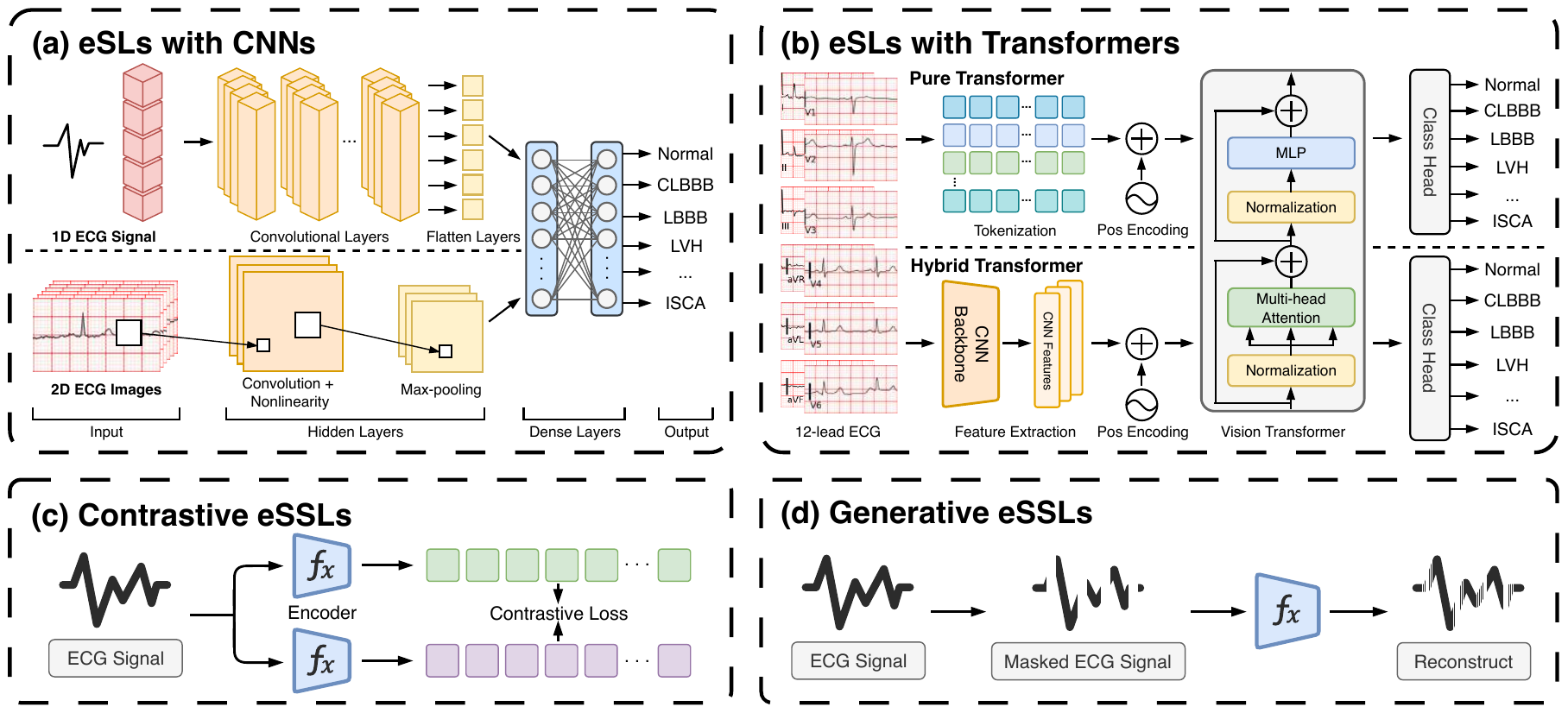}
    \vspace{-20pt}
    \caption{Current ECG representation learning methods, including (a) CNN-based supervised learning, (b) Transformer-based supervised learning, (c) contrastive learning, and (d) generative learning.}
    \label{fig:related}
    \vspace{-10pt}
\end{figure}

\section{Related Work}

\noindent \textbf{ECG Supervised Learning.} ECG supervised learning (eSL) methods, using CNNs or Transformers in Figure~\ref{fig:related}(a$-$b), achieve high accuracy in cardiovascular disease diagnosis. CNNs excel at capturing spatial and temporal patterns in 1D ECG signals or 2D ECG images \citep{tesfai2022lightweight, degirmenci2022arrhythmic, mashrur2019automatic, huang2022snippet}, while Transformers use attention mechanisms to model global dependencies \citep{natarajan2020wide, jiang2021hadln, he2023transformers}. Despite their strengths, eSLs rely heavily on large-scale datasets with expert-verified annotations, making them costly and impractical for pre-training tasks \citep{strodthoff2020deep}. This dependence limits their scalability and generalizability, particularly when addressing diverse datasets or unseen cardiac conditions.

\noindent \textbf{ECG Self-supervised Learning.} To overcome the annotation bottleneck, ECG self-supervised learning (eSSL) methods have been introduced, enabling representation learning from unannotated ECG signals in Figure~\ref{fig:related}(c$-$d). Contrastive learning frameworks, such as CLOCS and ASTCL \citep{kiyasseh2021clocs, wang2023adversarial}, explore temporal and spatial invariance in ECG data \citep{eldele2021time, chen2020simple, chen2021empirical}. Generative eSSL techniques reconstruct masked segments to capture signal-level features \citep{zhang2022maefe, sawano2022masked, na2024guiding, jinreading}. Despite their successes, eSSLs fail to incorporate clinical semantics from associated medical reports and require fine-tuning for downstream tasks \citep{liu2023improving, liu2023pixmim, he2022masked}, limiting their utility in zero-shot scenarios.

\noindent \textbf{ECG-Text Multimodal Learning.} Multimodal learning has advanced significantly in biomedical applications, especially in vision-language pre-training (VLP) frameworks for radiology \citep{liu2023g2d, liu2023m, wan2024med, zhang2023knowledge, wu2023medklip, abbaspourazad2023large}, which align radiology images with structured knowledge from reports to reduce noise and improve robustness. However, ECG-Text multimodal learning holds substantial potential for further development. Methods like MERL \citep{liu2024zero} and ECG-LM \citep{yangecg} integrate ECG signals and raw text reports but struggle with noise and inconsistencies in unstructured reports. Others, such as KED \citep{tian2024foundation}, use structured labels and contrastive learning strategies but face challenges from label noise and LLM-generated knowledge hallucinations. Our approach addresses these issues by structuring reports into meaningful entities, reducing noise, and aligning them with ECG signals without reliance on LLM-augmented content, minimizing hallucination risks while enabling efficient representation learning and downstream flexibility.


\begin{figure*}[htbp!]
    \hspace*{0.025\textwidth}
    \begin{minipage}[t]{0.55\textwidth}
        \subfigure[Design of ECG report entity extraction with (i) knowledge-enhanced prompt engineering, and (ii) candidate entity deduplication and mapping.]{
            \begin{minipage}{\textwidth}
                \centering
                \includegraphics[width=\textwidth]{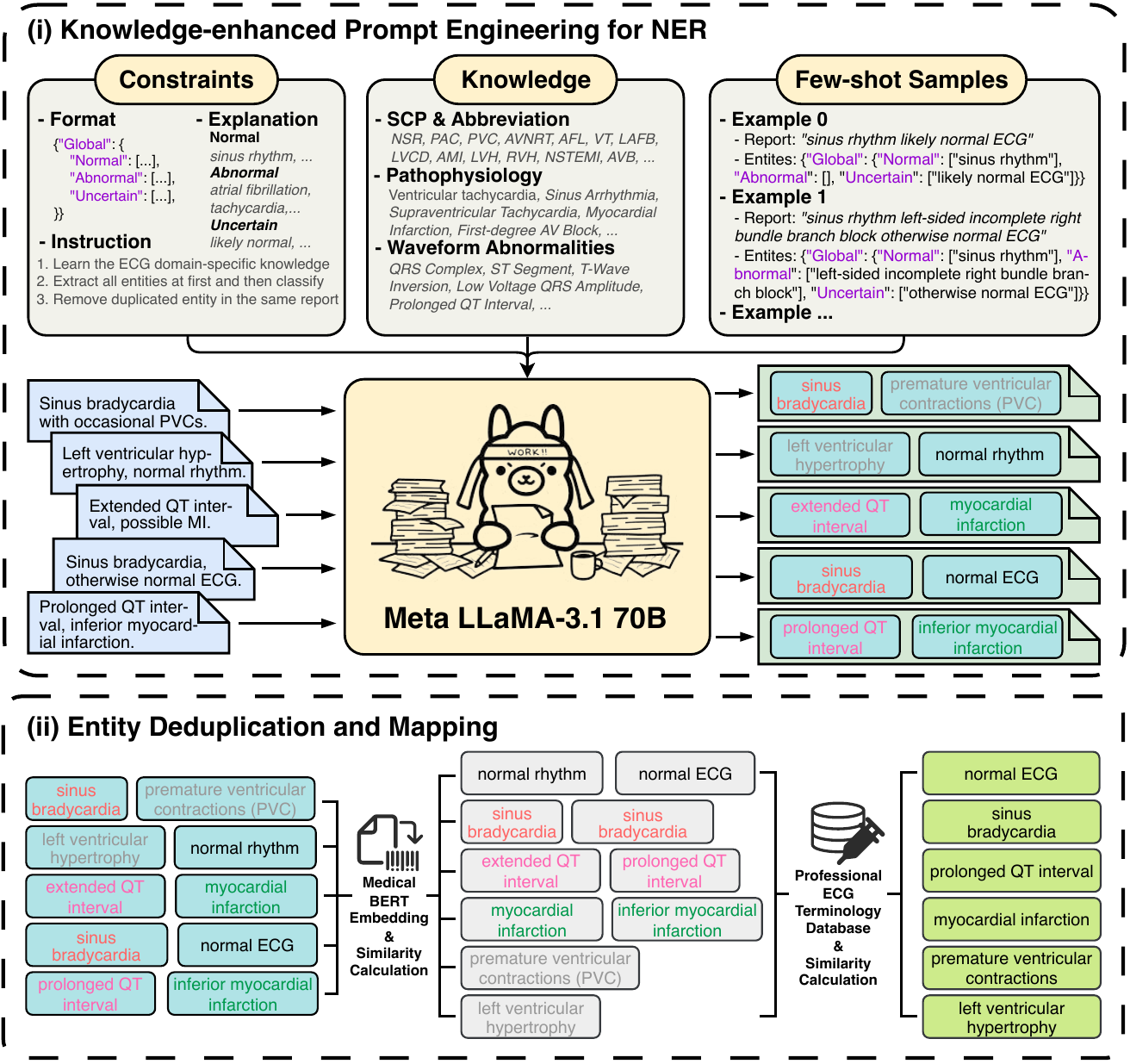}
            \end{minipage}
            \label{fig:ner4ecg}
        }
    \end{minipage}%
    \hspace*{0.03\textwidth}
    \begin{minipage}[t]{0.4\textwidth}
        \subfigure[ECG 1D ViT encoder in the SuPreME, with both lead-wise and position embedding.]{
            \includegraphics[width=0.85\textwidth]{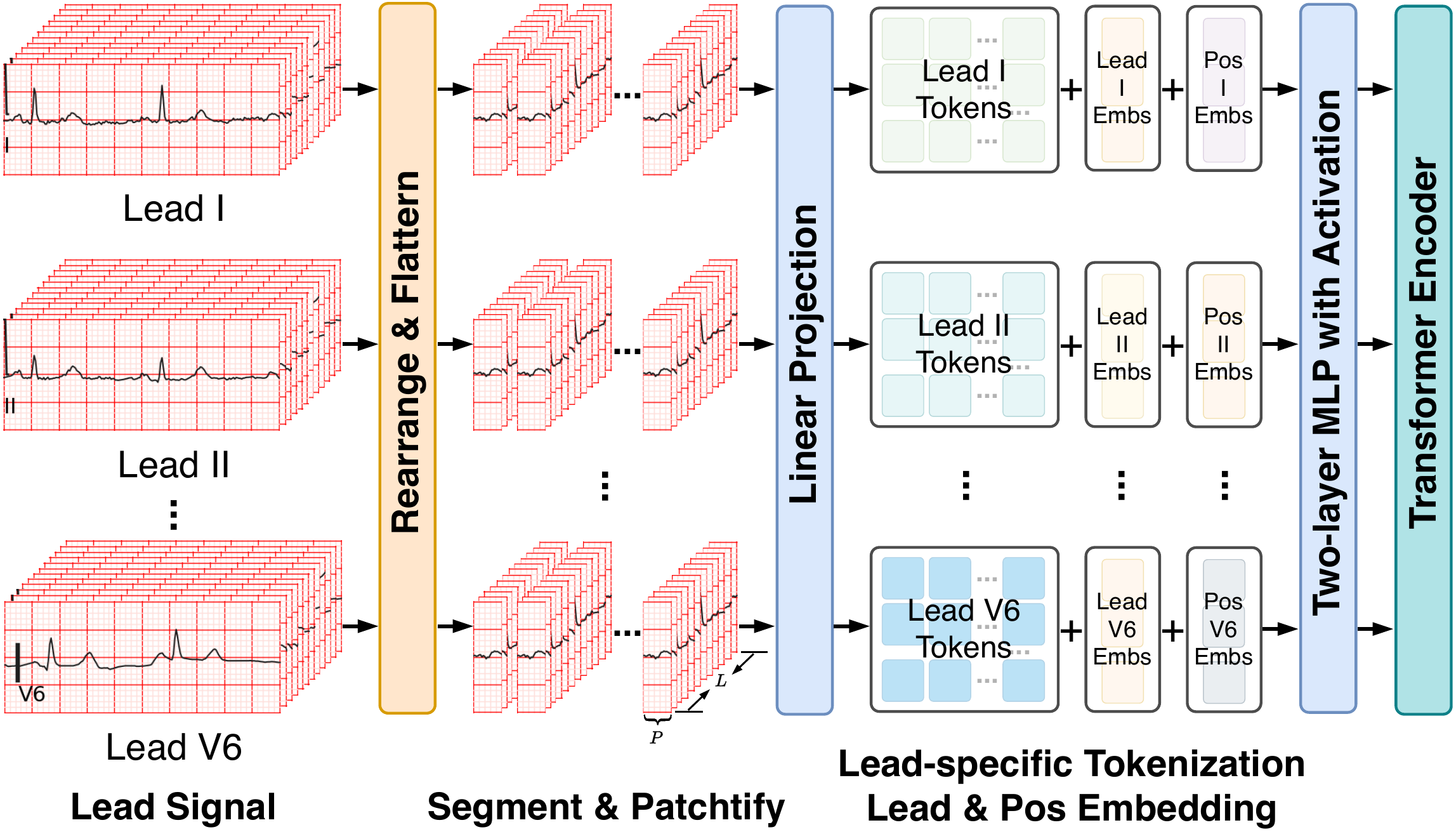}
            \label{fig:vit}
        }

        \vspace{-10pt}

        \subfigure[Architecture of the Cardiac Fusion Network (CFN) in the SuPreME.]{
                \includegraphics[width=0.85\textwidth]{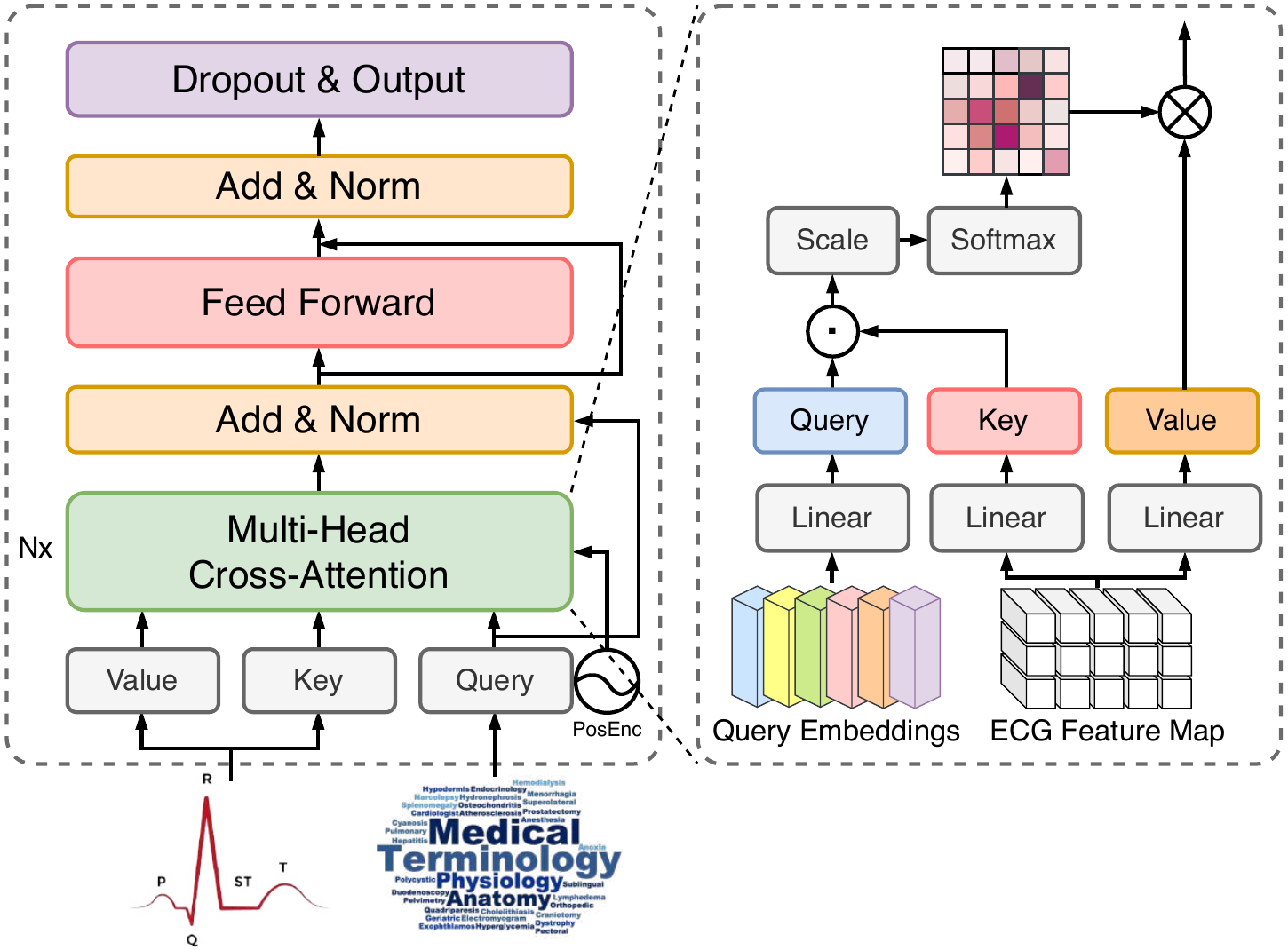}
            \label{fig:cfn}
        }
    \end{minipage}
    \vspace{-10pt}
    \caption{Implementation of the supervised ECG-Text multimodal pre-training framework including (a) ECG report entity extraction, (b) ECG 1D ViT encoder, and (c) architecture of the Cardiac Fusion Network.}
    \label{fig:method}
    \vspace{-15pt}
\end{figure*}

\section{Methodology}

SuPreME extracts structured clinical entities from ECG reports via an instruction-tuned LLM (Section~\ref{sec:ner4ecg}) to form cardiac queries, which are fused with ECG signals via Cardiac Fusion Network (CFN) in a shared latent space, enabling zero-shot classification of unseen cardiac conditions without fine-tuning (Section~\ref{sec:zero-shot}), thus yielding scalable, clinically meaningful representations.

\subsection{LLM-based Clinical Entity Extraction}
\label{sec:ner4ecg}

\noindent \textbf{Enrich LLM with Domain Knowledge.}
ECG reports generated by 12-lead devices (Appendix~\ref{app:ecg12lead}) contain diverse and nuanced descriptions of cardiac conditions. To contextualize these free-text reports for clinical entity extraction, we first construct a cardiac-specific vocabulary using GPT-4o-mini (Appendix~\ref{app:llm-ecg-ner-filter}), by filtering, normalizing, and aggregating terminology from clinician-validated resources such as SNOMED CT, UMLS, and SCP-ECG~\citep{bodenreider2004unified, donnelly2006snomed, rubel2016scp}. The vocabulary covers both complete diagnostic terms (e.g., \texttt{sinus rhythm}) and commonly used abbreviations (e.g., \texttt{LVH} for \texttt{Left Ventricular Hypertrophy}). This domain knowledge is then integrated into the LLM prompt design to guide the subsequent extraction process.

\noindent \textbf{Knowledge-Guided Entity Extraction.}
We employ an instruction-tuned LLM to extract clinical entities from unstructured ECG reports. The model is prompted using structured instructions and few-shot examples (Appendix~\ref{app:llm-ecg-ner-ext}), incorporating the curated cardiac vocabulary to enhance contextual understanding. It extracts diagnostic expressions (e.g., waveform patterns, cardiac abnormalities) along with their associated certainty. Extracted entities are categorized as \texttt{Normal}, \texttt{Abnormal}, or \texttt{Uncertain}, as illustrated in Figure~\ref{fig:ner4ecg}. Entities marked with low certainty (e.g., containing “probably” or “cannot rule out”) are discarded to improve diagnostic precision. Within each report, semantically similar expressions are merged to prepare for cross-report alignment and standardization.

\noindent \textbf{Entity Deduplication and Mapping.}
Although the extraction is guided and structured, lexical variability due to differing physician writing styles and device-specific formats still leads to redundant or inconsistent entities. To resolve this, we apply the curated cardiac vocabulary for entity standardization. Both extracted entities and reference terms are encoded using MedCPT, a medical BERT model pre-trained for clinical semantic similarity~\citep{jin2023medcpt}. Cosine similarity between embeddings is computed to perform soft matching. Entities that exhibit high average similarity to a reference term are aligned and deduplicated with a threshold selected and clinically validated by experienced cardiologists with over ten years of practice. Case study is provided in Appendix~\ref{app:llm-ecg-ner-case}.

\subsection{Multimodal ECG Supervised Learning}
\label{sec:spt4ecg}

\noindent \textbf{ECG Embedding with Vision Transformer.} 
The Vision Transformer (ViT) \citep{dosovitskiy2020image}, designed for 2D image processing, reshapes images into sequences of flattened patches for Transformer-based analysis. Similarly, ECG signals exhibit temporal and structural patterns analogous to the spatial relationships in images. We then adapt its architecture by dividing ECG time series into fixed-size patches, as shown in Figure~\ref{fig:vit}.

Multi-lead ECG signals are represented as $\mathbf{x} \in \mathbb{R}^{B \times L \times T}$, where $B$ is the batch size, $L$ the number of leads, and $T$ the number of time steps. Each lead's signal is independently segmented into $N = T / P$ non-overlapping patches of length $P$, resulting in $\mathbf{x}_{i,j} \in \mathbb{R}^{B \times P}$ for lead $i$ and patch index $j$. Each patch is flattened and passed through a shared linear projection layer $\mathbf{W}_p \in \mathbb{R}^{P \times D}$ to produce a token embedding $\mathbf{z}_{i,j} \in \mathbb{R}^{B \times D}$:

\vspace{-5pt}
\begin{equation}
    \begin{array}{c}
        \mathbf{z}_{i,j} = \mathbf{x}_{i,j} \mathbf{W}_p \\[0.3em]
        \mathbf{z'}_{i,j} = \mathbf{z}_{i,j} + \mathbf{e}_i + \mathbf{p}_j
    \end{array}
\end{equation}
\vspace{-5pt}

To preserve lead-specific features and spatial-temporal information, we introduce a unique learnable lead embedding $\mathbf{e}_i \in \mathbb{R}^D$ for each lead $i$, and a positional embedding $\mathbf{p}_j \in \mathbb{R}^D$ for each patch index $j$. These embeddings are element-wise added to the projected patch token $\mathbf{z}_{i,j}$, forming the enriched representation $\mathbf{z'}_{i,j}$. All enriched tokens from all leads and patches are then concatenated to form the Transformer encoder input sequence:

\vspace{-20pt}
\begin{equation}
    \begin{array}{c}
        \mathbf{Z} = [\mathbf{z'}_{1,1}, \dots, \mathbf{z'}_{1,N}, \dots, \mathbf{z'}_{L,N}] \\[0.3em]
        \mathbf{F}_{\text{ECG}} = \text{MLP}_{\text{ECG}}(\text{Dropout}(\mathbf{Z}))
    \end{array}
\end{equation}
\vspace{-13pt}

The final token sequence $\mathbf{Z} \in \mathbb{R}^{B \times (L \cdot N) \times D}$ is passed through Transformer encoders to extract high-level ECG representations. Each block consists of multi-head self-attention and feed-forward sublayers, with residual connections and layer normalization. To enhance generalization, we apply stochastic depth dropout to the residual paths.

Before multimodal fusion, the output features are passed through a modality-specific two-layer multilayer perceptron (MLP) projection head with an intermediate non-linear activation. This projection maps the ViT output from its internal width $D$ to a shared latent dimension $D'$ aligned with the textual modality, forming representation $\mathbf{F}_{\text{ECG}} \in \mathbb{R}^{B \times (L \cdot N) \times D'}$ used as the CFN input. Implementation details are provided in Appendix~\ref{app:implementation-trm}, ~\ref{app:implementation-ecg-proj}.

\noindent \textbf{Cardiac Query Embedding with MedCPT.}
Instead of relying on fixed categorical labels, our framework adopts a flexible and semantically meaningful approach based on textual cardiac queries. For the pre-training dataset, we construct fine-grained diagnostic queries by applying the LLM-based entity extraction pipeline described in Section~\ref{sec:ner4ecg}. Specifically, we generate a dataset-specific global query list of size $M$, consisting of standardized cardiac terms derived from the deduplicated and mapped output of the extraction process.

Let $\mathcal{Q} = \{q_1, q_2, \dots, q_M\}$ denote the global query list. Each query $q_i$ is encoded into a dense vector using the query encoder from MedCPT, which applies a Transformer (Trm) to the input sequence $[\text{CLS}] \ q_i \ [\text{SEP}]$. The final-layer $[\text{CLS}]$ token embedding is used as the query representation:

\vspace{-15pt}
\begin{equation}
    \begin{array}{c}
        \mathbf{E}[i, :] = \text{Trm}([\text{CLS}] \ q_i \ [\text{SEP}]) \\[0.3em]
        \mathbf{F}_{\text{Query}} = \text{MLP}_{\text{Query}}(\text{Dropout}(\mathbf{E}))
    \end{array}
\end{equation}
\vspace{-10pt}

All $M$ query embeddings $\mathbf{E} \in \mathbb{R}^{M \times 768}$ are then passed through a modality-specific two-layer MLP projection head with an intermediate activation function to obtain the final representations $\mathbf{F}_{\text{Query}} \in \mathbb{R}^{M \times D'}$ in the shared $D'$-dimensional latent space aligned with ECG token representations. Further implementations are provided in Appendix~\ref{app:implementation-text-proj}.

\noindent \textbf{Alignment by Cardiac Fusion Network.}
The Cardiac Fusion Network (CFN) fuses ECG signals with textual cardiac queries using a multi-layer Transformer decoder architecture, where query embeddings act as decoder inputs and ECG features serve as the encoder memory, following a standard cross-attention formulation (Figure~\ref{fig:cfn}).

Given a batch of ECG features $\mathbf{F}_{\text{ECG}} \in \mathbb{R}^{B \times (L \cdot N) \times D'}$ and query embeddings $\mathbf{F}_{\text{Query}} \in \mathbb{R}^{M \times D'}$, CFN fuses the two modalities through cross-attention. During pre-training, each ECG sample is paired with the same $M$ cardiac query embeddings, allowing CFN to learn a joint representation that captures query-conditioned patterns in the signal. The decoder attends to ECG patterns while grounding the prediction in the semantics of each diagnostic query, outputting $\mathbf{H} \in \mathbb{R}^{B \times M \times D'}$ which is passed through a single MLP classification head shared across all queries, producing $M$ binary logits per ECG sample, where each logit indicates the relevance of a specific query to the signal input:

\vspace{-10pt}
\begin{equation}
    \mathbf{Logits} = \text{MLP}_{\text{CFN}}(\mathbf{H}) \in \mathbb{R}^{B \times M}
\end{equation}
\vspace{-15pt}

In pre-training, we supervise CFN using weak binary labels derived from the entity extraction pipeline. Each ECG report is matched against the global query list of $M$ standardized diagnostic terms. A binary label of 1 is assigned if a mapped report entity matches a query; otherwise 0. This results in a sparse $M$-dimensional multi-label vector per ECG sample. To avoid data leakage and ensure modality separation, raw ECG reports are never used directly as input to the model. Instead, diagnostic query list serves as input with queries embedded independently and applied uniformly to every ECG sample. 

This formulation enables SuPreME to perform open-set classification with a flexible, scalable query interface, supporting multi-label learning while maintaining clear supervision-query decoupling. CFN initialization is in Appendix \ref{app:implementation-cfn-init}.

\subsection{Zero-shot Prompted Classification}
\label{sec:zero-shot}

To enable zero-shot classification on unseen cardiac conditions without fine-tuning, we construct concise, clinically meaningful prompts (e.g., \texttt{left bundle branch block} for \texttt{LBBB}) derived from SCP-ECG codes in each downstream dataset. These prompts form a dataset-specific query list aligned with the pre-training query space and serve as inputs to the textual modality of SuPreME.

We follow a simplified version of Clinical Knowledge-Enhanced Prompt Engineering (CKEPE)~\citep{liu2024zero}, where SCP-ECG codes are translated into discriminative phrases validated by UMLS and SNOMED CT. Unlike full CKEPE pipelines that retrieve verbose descriptions (e.g., \texttt{a condition characterized by prolonged QRS complex...} for \texttt{LBBB}), our approach promotes clarity and cross-modal fusion by avoiding redundant or overly detailed textual artifacts. These prompts are used exclusively during inference and remain fixed for all ECG samples within a dataset.

During zero-shot evaluation, ECG signals and textual prompts are encoded via the pre-trained encoders into $\mathbf{F}^{\text{eval}}_{\text{ECG}}$ and $\mathbf{F}^{\text{eval}}_{\text{Query}}$, then passed into the Cardiac Fusion Network (CFN). The CFN performs cross-modal attention to align features and outputs one logit per query-ECG pair. The final prediction scores are computed as:

\vspace{-18pt}
\begin{equation}
    \mathbf{Pred} = \sigma(\text{CFN}(\mathbf{F}^{\text{eval}}_\text{ECG}, \mathbf{F}^{\text{eval}}_\text{Query})) \in \mathbb{R}^{B \times M'}
\end{equation}
\vspace{-18pt}

This setup decouples the prediction space from any fixed label vocabulary, allowing the model to generalize to arbitrary diagnostic queries. The query list can vary across downstream datasets, and the classifier is query-agnostic, meaning no structural change is required when adapting to new tasks. Evaluation is conducted without fine-tuning using AUROC (AUC) per class and mean AUC across all prompts. Details about simplified-CKEPE, evaluation metrics are in Appendix~\ref{app:zeroshot-eval}.

\begin{figure*}[tp]
    \begin{minipage}[t]{0.45\textwidth}
        \subfigure{
            \begin{minipage}{\textwidth}
                \centering
                \resizebox{\textwidth}{!}{
                    \begin{tabular}{l|c|c|ccc}
                        \toprule[1.2pt]
                        & Evaluation & Zero-shot & \multicolumn{3}{c}{Linear Probing} \\
                        Framework & Approach & 0\% & 1\% & 10\% & 100\% \\
                        \midrule[1.2pt]
                        \multicolumn{6}{l}{\textbf{\textit{From Scratch}}} \\
                        \midrule
                        Random Init (CNN) & \textit{L} & - & 55.09 & 67.37 & 77.21 \\
                        Random Init (Transformer) & \textit{L} & - & 53.53 & 65.54 & 75.52 \\
                        \midrule
                        \multicolumn{6}{l}{\textbf{\textit{ECG Only}}} \\
                        \midrule
                        SimCLR \citep{chen2020simple} & \textit{L} & - & 58.24 & 66.71 & 72.82 \\
                        BYOL \citep{grill2020bootstrap} & \textit{L} & - & 55.78 & 70.61 & 74.92 \\
                        BarlowTwins \citep{zbontar2021barlow} & \textit{L} & - & 58.92 & 70.85 & 75.39 \\
                        MoCo-v3 \citep{chen2021empirical} & \textit{L} & - & 57.92 & 72.04 & 75.59 \\
                        SimSiam \citep{chen2021exploring} & \textit{L} & - & 59.46 & 69.32 & 75.33 \\
                        TS-TCC \citep{eldele2021time} & \textit{L} & - & 54.66 & 69.37 & 76.95 \\
                        CLOCS \citep{kiyasseh2021clocs} & \textit{L} & - & 56.67 & 70.91 & 75.86 \\
                        ASTCL \citep{wang2023adversarial} & \textit{L} & - & 57.53 & 71.15 & 75.98 \\
                        CRT \citep{zhang2023self} & \textit{L} & - & 56.62 & 72.03 & 76.65 \\
                        ST-MEM \citep{na2024guiding} & \textit{L} & - & 56.42 & 63.39 & 69.60 \\
                        \midrule
                        \multicolumn{6}{l}{\textbf{\textit{Multimodal Learning}}} \\
                        \midrule
                        MERL \citep{liu2024zero} & \textit{Z \& L} & \colorbox{gray!15}{73.54} & \textbf{63.57} & \textbf{78.35} & \colorbox{gray!15}{83.68} \\
                        \midrule
                        \textbf{SuPreME (Ours)} & \textit{Z \& L} & \textbf{77.20} & \colorbox{gray!15}{63.24} & \colorbox{gray!15}{72.34} & \textbf{84.48} \\
                        \bottomrule[1.2pt]
                    \end{tabular}
                }
                \vspace{-8pt}
                \captionof{table}{Performance of SuPreME and eSSLs, with '\textit{Z}' for zero-shot and '\textit{L}' for linear probing. Best results are \textbf{bolded} and second best \colorbox{gray!15}{gray}-flagged.}
                \label{tab:overall}
            \end{minipage}
        }


        \subfigure{
            \begin{minipage}{\textwidth}
                \centering
                \resizebox{\textwidth}{!}{
                    \begin{tabular}{l|cc|cc}
                        \toprule[1.2pt]
                        & \multicolumn{2}{c}{\textbf{\textit{Linear Classification}}} & \multicolumn{2}{c}{\textbf{\textit{Cardiac Fusion Network}}} \\
                        \midrule
                        Dataset & ResNet & ViT & ResNet & ViT \\
                        \midrule
                        PTB-XL-Superclass & 67.55 & 66.80 & 68.75 & \textbf{78.20} \\
                        PTB-XL-Subclass & 73.77 & 71.51 & 68.02 & \textbf{77.52}  \\
                        PTB-XL-Form & \textbf{64.34} & 62.10 & 58.85 & 60.67  \\
                        PTB-XL-Rhythm & 75.68 & 75.34 & 68.69 & \textbf{86.79}  \\
                        CPSC-2018 & \textbf{83.35} & 79.13 & 60.38 & 79.83 \\
                        CSN & 72.61 & 72.32 & 65.07 & \textbf{80.17} \\
                        \midrule
                        \textbf{Overall} & 72.88 & 71.23 & 64.96 & \textbf{77.20} \\
                        \bottomrule[1.2pt]
                    \end{tabular}
                }
                \vspace{-8pt}
                \captionof{table}{Performance of SuPreME and its variants on downstream datasets. Best results are \textbf{bolded}.}
                \label{tab:zero-shot}
            \end{minipage}
        }
    \end{minipage}%
    \hspace*{0.04\textwidth}
    \begin{minipage}[t]{0.5\textwidth}
        \subfigure{
            \includegraphics[width=\textwidth]{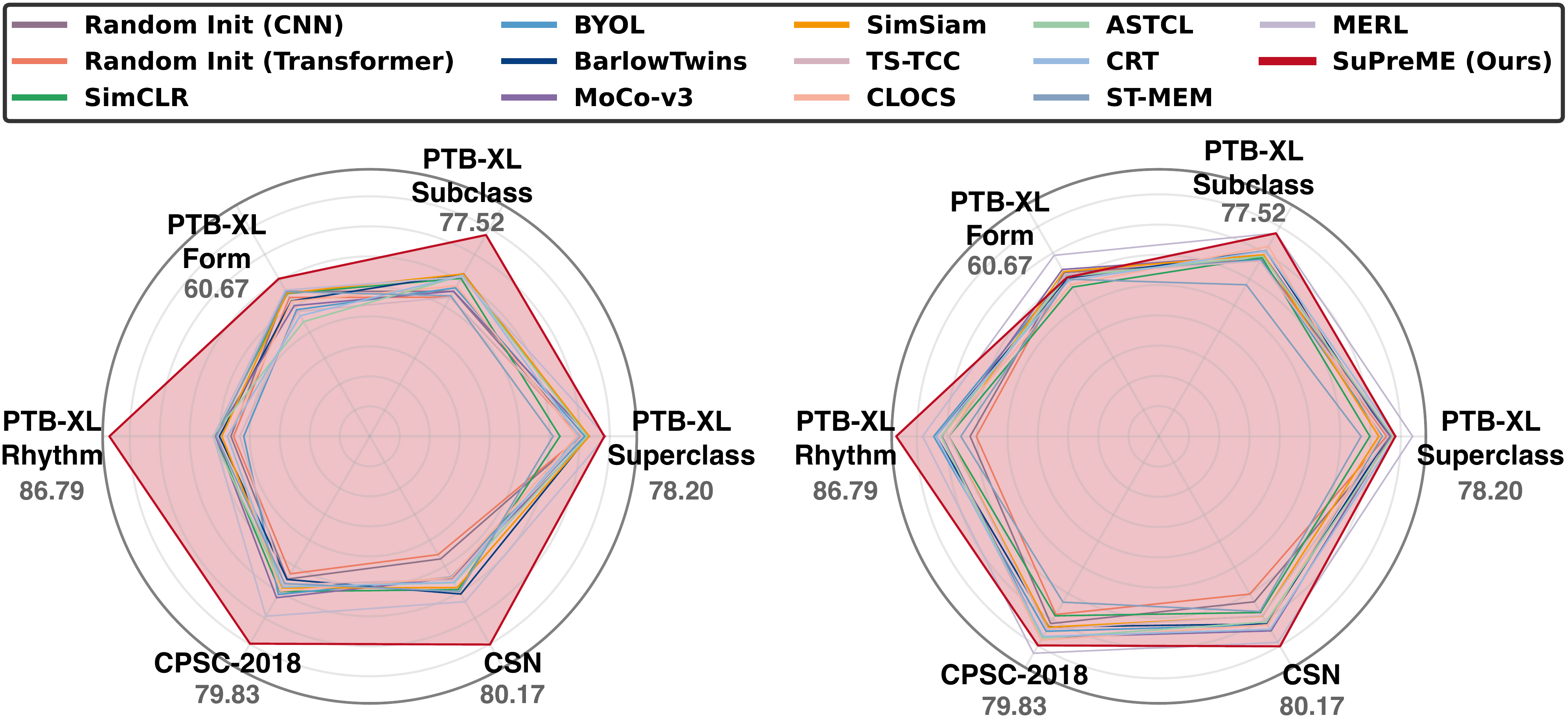}
        }
        \vspace{-15pt}
        \captionof{figure}{Comparison of SuPreME (zero-shot) and eSSLs (linear probing with 1\% data on the left and 10\% data on the right) across downstream datasets.}
        \label{fig:radar}

        \vspace{18.5pt}

        \subfigure{
            \includegraphics[width=\textwidth]{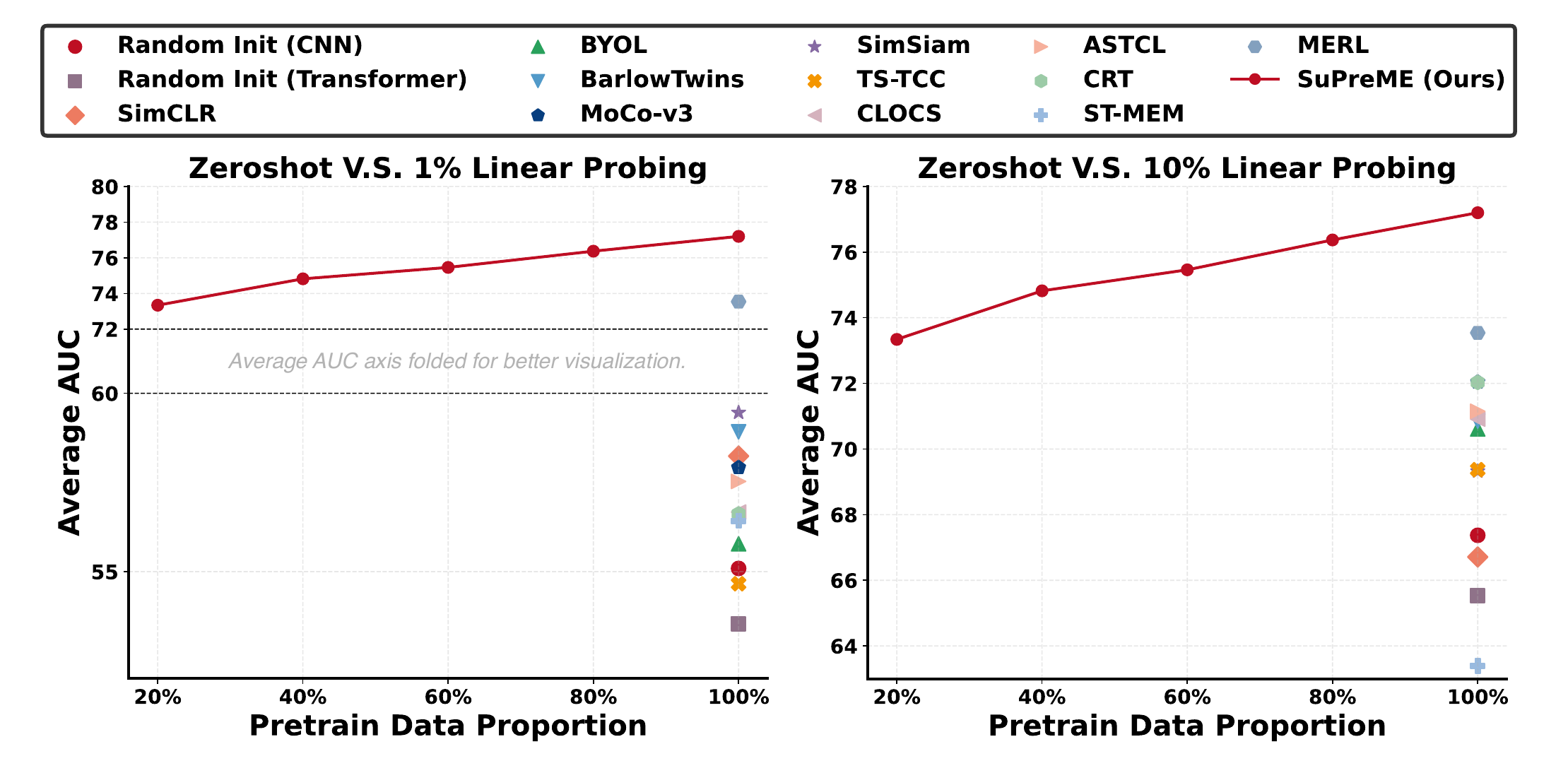}
        }
        \vspace{-15pt}
        \captionof{figure}{Data efficiency of SuPreME (zero-shot) and eSSLs (MERL in zero-shot, others in linear probing).}
        \label{fig:data}
    \end{minipage}
    \vspace{-15pt}
\end{figure*}

\begin{table*}[tb!]
\centering
\resizebox{\textwidth}{!}{
\begin{tabular}{lccc|ccc|ccc|ccc|ccc|ccc}
    \toprule[1.2pt]
     & \multicolumn{3}{c}{PTB-XL-Superclass} & \multicolumn{3}{c}{PTB-XL-Subclass} 
     & \multicolumn{3}{c}{PTB-XL-Form} & \multicolumn{3}{c}{PTB-XL-Rhythm} 
     & \multicolumn{3}{c}{CPSC-2018} & \multicolumn{3}{c}{CSN} \\
    Framework & 1\% & 10\% & 100\% & 1\% & 10\% & 100\% & 1\% & 10\% & 100\% 
           & 1\% & 10\% & 100\% & 1\% & 10\% & 100\% & 1\% & 10\% & 100\% \\
    \midrule[1.2pt]
    \textbf{\textit{From Scratch}} \\
    \midrule
    Random Init (CNN) & 70.45 & 77.09 & 81.61 & 55.82 & 67.60 & 77.91 & 55.82 & 62.54 & 73.00 & 46.26 & 62.36 & 79.29 & 54.96 & 71.47 & 78.33 & 47.22 & 63.17 & 73.13 \\
    Random Init (Transformer) & 70.31 & 75.27 & 77.54 & 53.56 & 67.56 & 77.43 & 53.47 & 61.84 & 72.08 & 45.36 & 60.33 & 77.26 & 52.93 & 68.00 & 77.44 & 45.55 & 60.23 & 71.37 \\
    \midrule
    \textbf{\textit{ECG Only}} \\
    \midrule
    SimCLR & 63.41 & 69.77 & 73.53 & 60.84 & 68.27 & 73.39 & 54.98 & 56.97 & 62.52 & 51.41 & 69.44 & 77.73 & 59.78 & 68.52 & 76.54 & 59.02 & 67.26 & 73.20 \\
    BYOL & 71.70 & 73.83 & 76.45 & 57.16 & 67.44 & 71.64 & 48.73 & 61.63 & 70.82 & 41.99 & 74.40 & 77.17 & 60.88 & 74.42 & 78.75 & 54.20 & 71.92 & 74.69 \\
    BarlowTwins & 72.87 & 75.96 & 78.41 & \colorbox{gray!15}{62.57} & 70.84 & 74.34 & 52.12 & 60.39 & 66.14 & 50.12 & 73.54 & 77.62 & 55.12 & 72.75 & 78.39 & 60.72 & 71.64 & 77.43 \\
    MoCo-v3 & 73.19 & 76.65 & 78.26 & 55.88 & 69.21 & 76.69 & 50.32 & \colorbox{gray!15}{63.71} & 71.31 & 51.38 & 71.66 & 74.33 & \colorbox{gray!15}{62.13} & 76.74 & 75.29 & 54.61 & 74.26 & 77.68 \\
    SimSiam & 73.15 & 72.70 & 75.63 & 62.52 & 69.31 & 76.38 & 55.16 & 62.91 & 71.31 & 49.30 & 69.47 & 75.92 & 58.35 & 72.89 & 75.31 & 58.25 & 68.61 & 77.41 \\
    TS-TCC & 70.73 & 75.88 & 78.91 & 53.54 & 66.98 & 77.87 & 48.04 & 61.79 & 71.18 & 43.34 & 69.48 & 78.23 & 57.07 & 73.62 & 78.72 & 55.26 & 68.48 & 76.79 \\
    CLOCS & 68.94 & 73.36 & 76.31 & 57.94 & 72.55 & 76.24 & 51.97 & 57.96 & 72.65 & 47.19 & 71.88 & 76.31 & 59.59 & \colorbox{gray!15}{77.78} & 77.49 & 54.38 & 71.93 & 76.13 \\
    ASTCL & 72.51 & 77.31 & 81.02 & 61.86 & 68.77 & 76.51 & 44.14 & 60.93 & 66.99 & \colorbox{gray!15}{52.38} & 71.98 & 76.05 & 57.90 & 77.01 & 79.51 & 56.40 & 70.87 & 75.79 \\
    CRT & 69.68 & 78.24 & 77.24 & 61.98 & 70.82 & 78.67 & 46.41 & 59.49 & 68.73 & 47.44 & 73.52 & 74.41 & 58.01 & 76.43 & 82.03 & 56.21 & 73.70 & 78.80 \\
    ST-MEM & 61.12 & 66.87 & 71.36 & 54.12 & 57.86 & 63.59 & 55.71 & 59.99 & 66.07 & 51.12 & 65.44 & 74.85 & 56.69 & 63.32 & 70.39 & 59.77 & 66.87 & 71.36 \\
    \midrule
    \textbf{\textit{Multimodal Learning}} \\
    \midrule
    MERL & \textbf{78.64} & \textbf{83.90} & \colorbox{gray!15}{85.27} & 61.41 & \textbf{77.55} & \colorbox{gray!15}{82.98} & \colorbox{gray!15}{56.32} & \textbf{69.11} & \textbf{77.66} & 52.16 & \textbf{78.07} & \colorbox{gray!15}{81.83} & \textbf{69.25} & \textbf{82.82} & \textbf{89.44} & \colorbox{gray!15}{63.66} & \textbf{78.67} & \colorbox{gray!15}{84.87} \\
    \midrule
    \textbf{SuPreME (Ours)} & \colorbox{gray!15}{73.58} & \colorbox{gray!15}{79.07} & \textbf{87.67} & \textbf{66.30} & \colorbox{gray!15}{74.20} & \textbf{84.84} & \textbf{58.94} & 58.93 & \colorbox{gray!15}{74.06} & \textbf{56.92} & \colorbox{gray!15}{76.27} & \textbf{84.42} & 58.28 & 70.51 & \colorbox{gray!15}{86.74} & \textbf{65.42} & \colorbox{gray!15}{75.08} & \textbf{89.16} \\
    \bottomrule[1.2pt]
\end{tabular}
}
\vspace{-5pt}
\caption{Specific linear probing performance of SuPreME and eSSLs across six downstream datasets. Best results are \textbf{bolded} and second best \colorbox{gray!15}{gray}-flagged.}
\label{tab:linearprobing}
\vspace{-15pt}
\end{table*}

\section{Experiments}

\subsection{Configuration and Settings}
\label{sec:config4exp}

\noindent \textbf{Clinical Entity Extraction.}
Following Section~\ref{sec:ner4ecg}, we extract and normalize clinical entities from MIMIC-IV-ECG using Llama3.1-70B-Instruct\footnote{Used offline for one-time inference only; not required during deployment. A large LLM ensures high-quality labels. (Appendix~\ref{app:dis-llm})} with structured prompts to ensure high-quality annotations. Entities are deduplicated via MedCPT embeddings (cosine similarity $> 0.8$) and mapped to UMLS/SNOMED CT (average cosine similarity $> 0.75$)\footnote{Verified by cardiologists with 10+ years of experience; Appendix~\ref{app:dis-threshold}, ~\ref{app:dis-llm-ner-eval}.}. Experiments are run on 8 NVIDIA A100-SMX4-80GB GPUs using vLLM~\cite{kwon2023efficient}. Statistics of extracted MIMIC-IV-ECG entities are in Appendix~\ref{app:mimic-stats}.

\noindent \textbf{Supervised ECG Pre-training.}
We use a 1D ViT-tiny encoder (patch size = 125, i.e., 0.25s) and a frozen MedCPT text encoder. Training employs AdamW (LR=$1\times10^{-3}$, weight decay=$1\times10^{-8}$) with cosine annealing ($T_0$=5000, $T_{\text{mult}}$=1, min LR=$1\times10^{-8}$), for up to 50 epochs with early stopping (patience=10, best AUC at 16). Batch size is set to 256 on 4 NVIDIA A100-PCIE-40GB GPUs\footnote{Compact pre-trained checkpoint (ViT-tiny + frozen MedCPT) runs on single GPU with $\geq$24GB memory, making it deployable in clinical or low-resource settings (Appendix~\ref{app:dis-compute}).}.

\noindent \textbf{Downstream Classification Task.}
SuPreME is evaluated on six unseen datasets (e.g., PTB-XL, CPSC-2018, and Chapman-Shaoxing-Ningbo) using dataset-specific prompts (Section~\ref{sec:zero-shot}). Ablation studies assess the impact of different ECG/text encoders and the CFN module as well as other key procedures. Mainstream eSSLs are benchmarked with linear probing by freezing ECG encoders and fine-tuning a linear layer on 1\%, 10\%, and 100\% of labeled data from the six datasets. All tasks are evaluated by average AUC across classes and datasets, following the data splits in Appendix \ref{app:datasplit}. Hyperparameters are provided in Appendix \ref{app:downstream} and overlap analysis in Appendix~\ref{app:overlap}.

\subsection{Evaluation with Mainstream eSSLs}

We evaluate SuPreME against mainstream eSSL frameworks across 106 classes in six downstream ECG datasets, conducting linear probing with eSSL ECG encoders across varying data proportions to facilitate performance comparison. Table~\ref{tab:overall} demonstrates AUC results of SuPreME and eSSLs under different evaluation approaches.

Our results demonstrate that SuPreME achieves superior performance compared to traditional eSSL frameworks. With an overall zero-shot AUC of 77.20\% (Details in Appendix~\ref{app:non-overlap}), SuPreME outperforms all non-multimodal eSSLs, which require linear probing even with 1\% (best: 59.46\%) or 10\% (best: 72.04\%) of labeled data, showcasing its strong generalization capabilities and efficient utilization of pre-trained knowledge. Even without the CFN module, SuPreME remains highly competitive (Table~\ref{tab:linearprobing}) using only the pre-trained ECG encoder for linear probing. Its overall performance consistently surpasses non-multimodal eSSL models across 1\%, 10\%, and 100\% labeled data (also outperforms in 15/18 tasks with different datasets and labelled data portion), and achieves comparable performance to multimodal contrastive learning frameworks like MERL (ViT backbone with explicit contrastive objectives, more comparisons in Appendix~\ref{app:dis-merl}).

Since SuPreME’s ECG backbone is optimized jointly with the CFN rather than via an explicit contrastive loss, part of the diagnostic knowledge is encoded within cross-modal interactions. Linear probing on the ECG encoder alone thus cannot fully utilize the rich alignment captured during pre-training. This explains why the full SuPreME pipeline including query prompts and CFN achieves stronger zero-shot performance, even compared to linear probing with more labeled data. We highlight zero-shot performance as our main evaluation objective, as it aligns with real-world clinical settings where labeled ECG data is scarce and fine-tuning is often impractical.

Figure~\ref{fig:radar} presents framework performance across individual datasets. SuPreME's advantage on the PTB-XL-Superclass dataset is minimal, likely due to the dataset's simplicity, as it includes only 5 broad cardiac condition labels (e.g., \texttt{NORM}, \texttt{STTC}, \texttt{MI}), making it difficult to differentiate model performance. All frameworks perform poorly on the PTB-XL-Form dataset, which focuses on 19 ECG waveform types that do not directly correspond to cardiac conditions, leading to ambiguous associations and reduced performance for all models.

To investigate SuPreME’s sensitivity to pre-training scale, we evaluate its zero-shot performance under varying data proportions (Figure~\ref{fig:data}). SuPreME consistently improves with more data and maintains a clear advantage over non-multimodal eSSLs. Remarkably, with only 20\% of pre-training data, SuPreME outperforms all non-multimodal eSSLs using 1\% or 10\% labeled data for linear probing, and matches the zero-shot performance of the multimodal baseline MERL trained on 100\% of the pre-training data. Moreover, SuPreME’s zero-shot performance with just 20\% of data also exceeds MERL’s linear probing result with 1\% labels, highlighting its superior generalization and efficiency under limited supervision. Notably, SuPreME achieves these results with significantly fewer computational resources and shorter training time\footnote{SuPreME: 4$\times$A100-40GB GPUs for 16 epochs ($\sim$90 minutes); MERL: 8$\times$A100-40GB GPUs for 50 epochs ($\geq$1 day).} (Appendix~\ref{app:dis-compute}).

\begin{figure*}[ht!]
    \hspace*{0.02\textwidth}
    \begin{minipage}[t]{0.395\textwidth}
        \subfigure{
            \includegraphics[width=\textwidth]{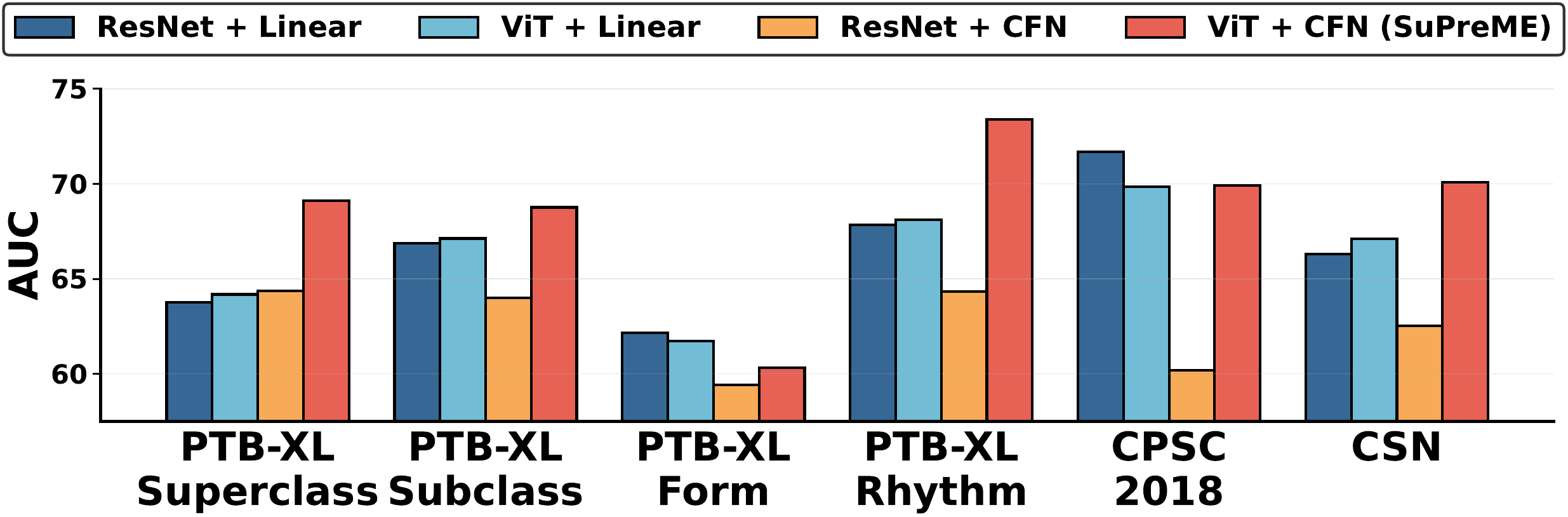}
        }
        \vspace{-20pt}
        \captionof{figure}{Specific zero-shot performance of SuPreME and its variants across downstream datasets.}
        \label{fig:variant}


        \subfigure{
            \includegraphics[width=\textwidth]{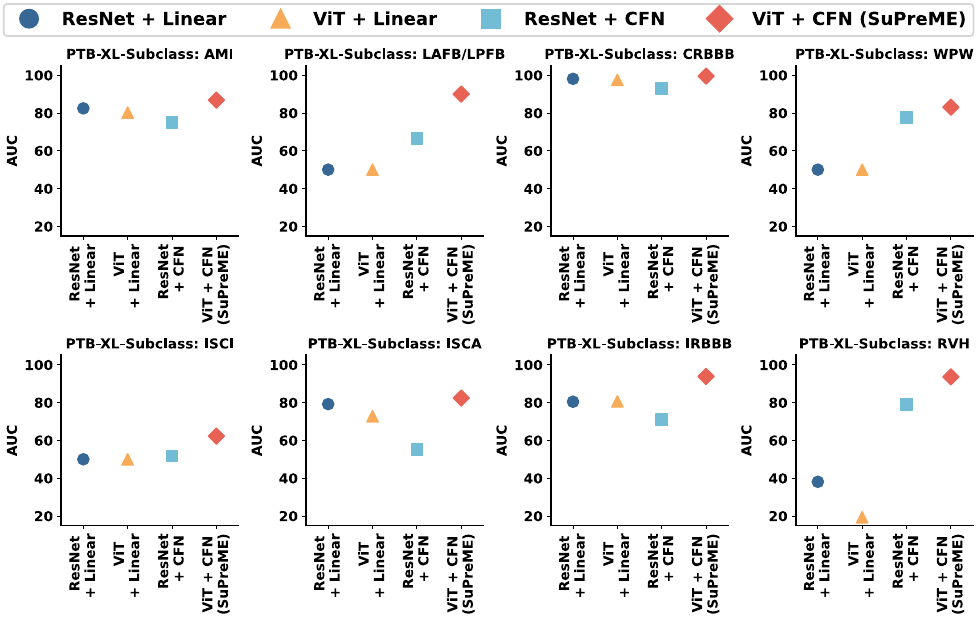}
        }
        \vspace{-20pt}
        \captionof{figure}{Specific zero-shot classification AUC performance of SuPreME and its variants on selected detailed categories in PTB-XL-Subclass.}
        \label{fig:scatter_ptbxl_subclass_selected}
    \end{minipage}
    \hspace*{0.02\textwidth}
    \begin{minipage}[t]{0.5\textwidth}
        \centering
        \begin{minipage}[t]{0.49\textwidth}
            \subfigure{
                \resizebox{1.13\textwidth}{!}{
                    \begin{tabular}{c|c}
                        \toprule[1.2pt]
                        LLM Size & Zero-shot AUC \\
                        \midrule[1.2pt]
                        Llama3.1-8B-Instruct & 72.89 $\pm$ 0.49 \\
                        \midrule
                        \textbf{Llama3.1-70B-Instruct (Ours)} & \textbf{77.20 $\pm$ 0.21} \\
                        \bottomrule[1.2pt]
                    \end{tabular}
                }
            }
            \vspace{-11.5pt}
            \captionof{table}{LLM for entity extraction.}
            \label{tab:llmner}
        \end{minipage}
        \hfill
        \begin{minipage}[t]{0.49\textwidth}
            \subfigure{
                \resizebox{0.9\textwidth}{!}{
                    \begin{tabular}{c|c}
                        \toprule[1.2pt]
                        Uncertainty Filtering & Zero-shot AUC \\
                        \midrule[1.2pt]
                        Not Filtered & 63.96 $\pm$ 0.47 \\
                        \midrule
                        \textbf{Filtered (Ours)} & \textbf{77.20 $\pm$ 0.21} \\
                        \bottomrule[1.2pt]
                    \end{tabular}
                }
            }
            \vspace{-11pt}
            \captionof{table}{Uncertain entity filtering.}
            \label{tab:ettfilter}
        \end{minipage}
        \hspace*{-0.06\textwidth} 
    

        \hspace*{0.05\textwidth} 
        \begin{minipage}[t]{0.45\textwidth}
            \subfigure{
                \resizebox{\textwidth}{!}{
                    \begin{tabular}{c|c}
                        \toprule[1.2pt]
                        Deduplication & Zero-shot AUC \\
                        \midrule[1.2pt]
                        Not Deduplicated & 65.94 $\pm$ 0.49 \\
                        \midrule
                        \textbf{Deduplicated (Ours)} & \textbf{77.20 $\pm$ 0.21} \\
                        \bottomrule[1.2pt]
                    \end{tabular}
                }
            }
            \vspace{-12pt}
            \captionof{table}{Entity deduplication.}
            \label{tab:dedup}
        \end{minipage}
        \hfill
        \hspace*{-0.03\textwidth} 
        \begin{minipage}[t]{0.42\textwidth}
            \subfigure{
                \resizebox{0.83\textwidth}{!}{
                    \begin{tabular}{c|c}
                        \toprule[1.2pt]
                        Backbone & Zero-shot AUC \\
                        \midrule[1.2pt]
                        ResNet & 64.96 $\pm$ 0.20 \\
                        \midrule
                        \textbf{ViT (Ours)} & \textbf{77.20 $\pm$ 0.21} \\
                        \bottomrule[1.2pt]
                    \end{tabular}
                }
            }
            \vspace{-11pt}
            \captionof{table}{ECG backbone\\encoders.}
            \label{tab:backbone}
        \end{minipage}
    

        \hspace*{0.06\textwidth}
        \begin{minipage}[t]{0.49\textwidth}
            \subfigure{
                \resizebox{0.85\textwidth}{!}{
                    \begin{tabular}{c|c}
                        \toprule[1.2pt]
                        Language Model & Zero-shot AUC \\
                        \midrule[1.2pt]
                        BioClinicalBERT & 62.95 $\pm$ 0.53 \\
                        \midrule
                        PubMedBERT & 62.51 $\pm$ 2.21 \\
                        \midrule
                        \textbf{MedCPT (Ours)} & \textbf{77.20 $\pm$ 0.21} \\
                        \bottomrule[1.2pt]
                    \end{tabular}
                }
            }
            \vspace{-11pt}
            \captionof{table}{Language model\\encoders.}
            \label{tab:lm}
        \end{minipage}
        \hfill
        \hspace*{-0.04\textwidth}
        \begin{minipage}[t]{0.45\textwidth}
            \vspace{10.5pt}
            \subfigure{
                \resizebox{0.92\textwidth}{!}{
                    \begin{tabular}{c|c}
                        \toprule[1.2pt]
                        Module & Zero-shot AUC \\
                        \midrule[1.2pt]
                        w/o CFN (Linear) & 72.70 $\pm$ 0.42 \\
                        \midrule
                        \textbf{CFN (Ours)} & \textbf{77.20 $\pm$ 0.21} \\
                        \bottomrule[1.2pt]
                    \end{tabular}
                }
            }
            \vspace{-11pt}
            \captionof{table}{Cardiac Fusion\\Network (CFN).}
            \label{tab:cfn}
        \end{minipage}

        
        \hspace*{0.04\textwidth}
        \begin{minipage}[t]{0.49\textwidth}
            \subfigure{
                \resizebox{\textwidth}{!}{
                    \begin{tabular}{c|c}
                        \toprule[1.2pt]
                        Prompt Strategy & Zero-shot AUC \\
                        \midrule[1.2pt]
                        GPT-4o Generated & 60.83 $\pm$ 0.26 \\
                        \midrule
                        CKEPE Detailed & 69.16 $\pm$ 1.94 \\
                        \midrule
                        \textbf{CKEPE Simplified (Ours)} & \textbf{77.20 $\pm$ 0.21} \\
                        \bottomrule[1.2pt]
                    \end{tabular}
                }
            }
            \vspace{-12pt}
            \captionof{table}{Zero-shot cardiac query prompts.}
            \label{tab:ckepe}
        \end{minipage}
        \hfill
        \hspace*{-0.02\textwidth}
        \begin{minipage}[t]{0.4\textwidth}
            \subfigure{
                \resizebox{0.9\textwidth}{!}{
                    \begin{tabular}{c|c}
                        \toprule[1.2pt]
                        Dropout Ratio & Zero-shot AUC \\
                        \midrule[1.2pt]
                        0.05 & 75.98 $\pm$ 0.56 \\
                        \midrule
                        \textbf{0.10 (Ours)} & \textbf{77.20 $\pm$ 0.21} \\
                        \midrule
                        0.15 & 75.63 $\pm$ 0.63 \\
                        \bottomrule[1.2pt]
                    \end{tabular}
                }
            }
            \vspace{-11.4pt}
            \captionof{table}{Pre-training\\dropout ratios.}
            \label{tab:dropout}
        \end{minipage}
        
    \end{minipage}
    \vspace{-15pt}
\end{figure*}

\subsection{Evaluation of SuPreME Architecture}

Beyond comparisons with eSSLs, we assess the contribution of core components in SuPreME by varying its core modules, including the ECG backbone (ResNet vs. ViT) and (Linear vs. CFN) shown in Table~\ref{tab:zero-shot} and Figure~\ref{fig:variant}. Overall, SuPreME (ViT + CFN) achieves the highest average AUC of 77.20\%, with strong results on PTB-XL-Rhythm (86.79\%) and CSN (80.17\%), demonstrating the effectiveness of cross-modal fusion for temporally and spatially complex signals.

Under linear classification, ResNet outperforms ViT across most datasets, reflecting its inductive bias toward local feature extraction. However, once CFN is introduced, ViT significantly benefits from its attention mechanisms and structured prompts, outperforming all other variants. This suggests that ViT’s global receptive field aligns well with the query-driven fusion in CFN, while ResNet’s local filters are less suited for attending over sparse textual queries.

Notably, the performance of ResNet + CFN is lower than ResNet + Linear across several datasets. We attribute this to a mismatch between ResNet’s hierarchical, spatially localized features and CFN’s attention-based fusion, which benefits more from globally contextualized inputs like ViT. CFN is designed to interpret semantically aligned queries over long-range dependencies, an area where ResNet lacks representational flexibility. This highlights the importance of matching the backbone’s encoding characteristics with the fusion strategy. Details are in Appendix~\ref{app:dis-cfn-backbone}.

Figure~\ref{fig:scatter_ptbxl_subclass_selected} further analyzes SuPreME’s performance on individual cardiac conditions in PTB-XL-Subclass (Complete results in Appendix~\ref{app:scatters}). SuPreME consistently achieves high AUCs (many $>90$), especially for nuanced conditions like LAFB/LPFB, CRBBB, CLBBB, and RVH, where both query semantics and signal patterns must be integrated. In contrast, ResNet + CFN underperforms in complex arrhythmias (e.g., ISCA, IRBBB), reinforcing our insight that strong multimodal fusion requires compatible encoders.



Unlike linear classifiers with fixed output dimensions, CFN enables flexible prompt-driven classification, aligning query semantics with signal patterns in a shared latent space as specified in Section~\ref{sec:spt4ecg}, improving generalization to novel conditions without fine-tuning and demonstrates its strength especially when paired with ViT.

\subsection{Ablation Analysis}
\label{sec:ablation}

\noindent \textbf{Entity Extraction Model.}
Using Llama3.1-70B-Instruct for NER improves zero-shot AUC by 4.31\% over the 8B variant (Table~\ref{tab:llmner}), reflecting better supervision quality. This step is offline and serves to build a high-quality labeled dataset, not for deployment.

\noindent \textbf{Uncertain Entity Filtering.}  
A cardiac-specific vocabulary with synonym consolidation ensures reliable alignment (Table~\ref{tab:query-case}). Removing the uncertainty filtering step lowers zero-shot AUC from 77.20\% to 63.96\% (Table~\ref{tab:ettfilter}), confirming its importance.

\noindent \textbf{Clinical Entity Mapping.}
Mapping to a standardized 295-term vocabulary improves zero-shot AUC from 65.94\% to 77.20\% (Table~\ref{tab:dedup}), likely by removing noise and resolving label redundancy to better represent distinct cardiac conditions.

\noindent \textbf{ECG Encoder Backbone.}  
Replacing ViT-tiny with ResNet18 drops zero-shot AUC by 12.24\% (Table~\ref{tab:backbone}), suggesting ResNet is less effective at modeling long-range ECG dependencies than ViT (Appendix~\ref{app:dis-cfn-backbone}).

\noindent \textbf{Clinical Text Encoder.}
Among BioClinicalBERT, PubMedBERT, and MedCPT, the latter achieves the highest AUC, outperforming the others by over 14.25\% (Table~\ref{tab:lm}), likely due to its contrastive training objective, which better captures fine-grained clinical distinctions.

\noindent \textbf{Cardiac Fusion Network Module.}
We compare CFN-based fusion with a simple linear projection (Table~\ref{tab:cfn}). CFN lifts the zero-shot AUC from 72.70\% to 77.20\%, highlighting the benefits of cross-attention in capturing multimodal synergies between ECG signals and text queries.

\noindent \textbf{Customized Cardiac Prompts.}
Among three strategies (GPT-4o, detailed CKEPE, and simplified CKEPE), the simplified CKEPE achieves the best AUC (Table~\ref{tab:ckepe}), with $\geq$ 8.04\% improvement, suggesting that concise, clinically focused prompts enhance alignment and reduce noise.

\noindent \textbf{Dropout Ratio.}
In pre-training, we compare dropout rates of \{0.05, 0.10, 0.15\}. A rate of 0.10 yields the best AUC, striking a balance between regularization and signal retention (Table~\ref{tab:dropout}).

\section{Conclusion}

We present a novel LLM-based method for ECG clinical entity extraction and construct a high-quality labeled dataset from MIMIC-IV-ECG. Building on this, we propose SuPreME, a scalable supervised pre-training framework for multimodal ECG representation learning that fuses ECG signals with fine-grained, standardized medical terminologies rather than free-text reports. Its Cardiac Fusion Network (CFN) and simplified Clinical Knowledge-Enhanced Prompt Engineering (CKEPE) eliminate the need for further fine-tuning, enabling robust zero-shot classification with concise cardiac queries.
Benchmarked on six downstream datasets, SuPreME achieves superior zero-shot performance against 11 eSSLs, underscoring both data efficiency and diagnostic precision. Our results highlight the value of explicit entity-level supervision over raw text alignment in ECG multimodal learning, providing a strong basis for clinically oriented ECG representation learning.

\section*{Limitations}

While SuPreME achieves strong zero-shot performance, several limitations remain. First, the clinical entity extraction relies on a large language model not fine-tuned for cardiology, which may miss rare or ambiguous terms and introduce noise. Second, SuPreME assumes generalization across clinical settings, but real-world data often involve device variability, demographic shifts, and class imbalance. Our experiments show lower performance on rare conditions, indicating sensitivity to distribution shift. Additionally, because the ECG encoder is trained jointly with CFN rather than via contrastive objectives, its features alone may not always outperform other baselines under linear probing. Lastly, most existing ECG baselines are single-modal and few of them support open zero-shot evaluation (e.g., MERL), underscoring the need for clinically motivated zero-shot benchmarks that better reflect practical deployment scenarios and support fairer comparison across methods.

\clearpage

\bibliography{main}

\clearpage

\appendix

\section{Appendix}

\subsection{Dataset and Model Overview}
\label{app:data-model}

\subsubsection{Pre-training Dataset}
\label{app:data-model-pretrain}

\textbf{MIMIC-IV-ECG.}
MIMIC-IV-ECG\footnote{MIMIC-IV-ECG is available at \href{https://physionet.org/content/mimic-iv-ecg/1.0/}{https://physionet.org/content/mimic-iv-ecg/1.0/}.} is a comprehensive database containing 800,035 diagnostic ECG samples from 161,352 unique patients, with 12-lead recordings in 10 second length and sampled at 500 Hz~\cite{gow2023mimic}. These data have been matched with patient records in the MIMIC-IV clinical database, allowing for the association of waveforms with reports when a cardiologist’s report is available through provided linking information. To enhance the usability of the data, we exclude empty reports as well as reports containing fewer than 3 words, and replace 'NaN' and 'Inf' values in the ECG records with the average of 6 neighboring points. Ultimately, the dataset used for clinical entity extraction tasks includes 771,500 samples, each comprising 18 machine-generated ECG reports based on rules and the corresponding ECG data. After clinical NER and deduplication on the 18 ECG reports of each sample, the dataset holds 295 labels of professional medical terminologies.

\subsubsection{Downstream Dataset}
\label{app:data-model-downstream}

\textbf{PTB-XL.}
PTB-XL\footnote{PTB-XL is available at \href{https://physionet.org/content/ptb-xl/1.0.3/}{https://physionet.org/content/ptb-xl/1.0.3/}.} is a large open-source ECG dataset, comprising 21,799 clinical ECG records from 18,869 patients, with each lead sampled at a rate of 500 Hz and a duration of 10 seconds~\cite{wagner2020ptb}. A total of 71 different ECG reports are SCP-ECG compliant, covering diagnostic, form and rhythm reports.
PTB-XL also provides a recommended train-test split and includes multi-level ECG annotations, covering Superclass (5 categories), Subclass (23 categories), Form (19 categories), and Rhythm (12 categories). Notably the 4 subsets have different sample sizes.

\noindent \textbf{CPSC-2018.} 
The CPSC-2018\footnote{CPSC-2018 is available at \href{http://2018.icbeb.org/Challenge.html}{http://2018.icbeb.org/Challenge.html}.} dataset originates from the China Physiological Signal Challenge (CPSC) 2018, including 6,877 records from 9,458 patients, with durations ranging from 6 to 60 seconds~\cite{liu2018open}. The standard 12-lead ECG data is sampled at a rate of 500 Hz, collected from 11 hospitals and categorized into 9 different labels: 1 normal type and 8 abnormal types.

\noindent \textbf{Chapman-Shaoxing-Ningbo (CSN).}
The CSN\footnote{Chapman-Shaoxing-Ningbo is available at \href{https://physionet.org/content/ecg-arrhythmia/1.0.0/}{https://physionet.org/content/ecg-arrhythmia/1.0.0/}.} 12-lead ECG dataset is created with the support of Chapman University, Shaoxing People's Hospital and Ningbo First Hospital, which includes 12-lead ECGs from 45,152 patients, with a sampling rate of 500 Hz and a duration of 10 seconds~\cite{zheng2020optimal, zheng2022large}. It contains expert annotated features that cover variety of common heart rhythms and other cardiovascular conditions. We exclude ECG records with "unknown" annotations and get 23,026 ECG records with 38 different labels.

\subsubsection{Llama3.1-70B-Instruct Model}
\label{app:data-model-llama}

Llama3.1-70B-Instruct\footnote{Llama3.1-70B-Instruct is available at \href{https://huggingface.co/meta-llama/Llama-3.1-70B-Instruct}{https://huggingface.co/meta-llama/Llama-3.1-70B-Instruct}.} is a 70-billion parameter large language model released by Meta AI as part of the Llama 3 family. Built on a transformer decoder architecture, it is optimized for instruction following and few-shot generalization through extensive supervised fine-tuning and reinforcement learning from human feedback (RLHF). Compared to its predecessors, Llama3.1-70B-Instruct demonstrates substantial improvements in reasoning, factuality, and alignment with user intent across a wide range of NLP tasks.

In our framework, we leverage Llama3.1-70B-Instruct to extract fine-grained diagnostic entities from free-text ECG reports in the MIMIC-IV-ECG dataset. The scale and instruction-tuning of this model make it well suited for domain-specific named entity recognition (NER) in noisy clinical narratives. Our objective is to construct a high-quality, large-scale set of cardiac entities and their mapped terminologies, enabling robust supervision for ECG-text multimodal learning and promoting reproducibility in future research.

Although smaller models can provide acceptable results (see Section~\ref{sec:ablation}), we adopt Llama3.1-70B-Instruct to maximize annotation quality, particularly for downstream applications in clinical and low-resource settings that rely on precise structured supervision.

\subsection{Electrocardiogram (ECG)}
\label{app:ecg12lead}

In the medical field, electrocardiogram (ECG) is an important tool for recording and analyzing patients' cardiac activities, which helps healthcare professionals identify various kinds of cardiac problems by detecting the electrical changes in different leads. The standard 12-lead ECG is the most common method of recording ECGs, and it can capture relatively comprehensive range of cardiac signals through placing electrodes at different locations on the body, providing information of the heart's health conditions.

\begin{figure}[h]
    \includegraphics[width=\columnwidth]{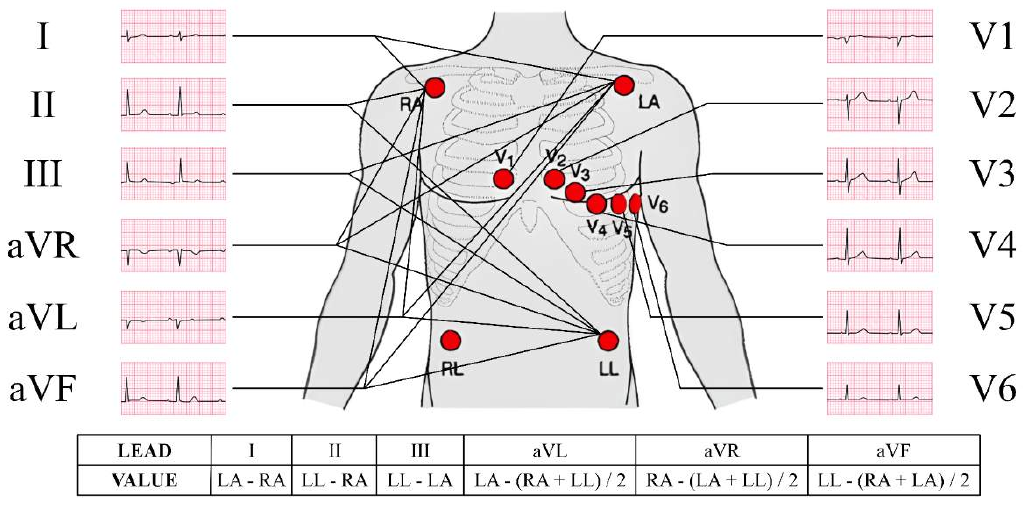}
    \caption{Standard 12-lead Electrocardiogram (ECG) showing 'sinus rhythm'.}
    \label{fig:ecg12lead}
    \vspace{-20pt}
\end{figure}

The basic components of the 12-lead ECG include the limb leads and the precordial leads. The limb leads contain I, II, III, aVR, aVL, and aVF, each of them consists of a combination of electrodes located primarily in the right arm, left arm, left leg, and right leg (as shown in Figure~\ref{fig:ecg12lead}). The precordial leads contain V1, V2, V3, V4, V5, and V6, which all correspond to specific single electrodes at different locations on the chest, and are used to observe in detail the electrical activity of the anterior, lateral, and posterior walls of the heart.

\subsection{Clinical NER Prompts, Statistics and Case Study}
\label{app:llm-ecg-ner}

\subsubsection{Prompt for Medical Database Terminology Filtering}
\label{app:llm-ecg-ner-filter}

\begin{lstlisting}[language=Python, numbers=none]
system_message = """\
You are a clinical NLP assistant specializing in identifying ECG related terminologies from medical databases.

Your primary task is to serve as a strict terminology filter that judges whether the provided terminology is related to ECG or not, and output your judgement in a **strictly formatted JSON object** that conforms exactly to the following schema:

{
  "IS_ECG_TERM": true/false
}

**Strict constraints**:
- Return **only** the JSON object. Do not include any natural language explanation or commentary.
- Do not hallucinate or invent fields not specified above.

Your output will be used in real-life clinical settings. Any deviation from this format may cause serious issues in downstream applications. Be precise and compliant.
"""

def get_prompt(row):
    return f"""\
Please read and give your judgement on the following terminology.

Terminology:
\"{row["ENG_TERM"]}\"
\end{lstlisting}

\subsubsection{Prompt for Report Entity Extraction}
\label{app:llm-ecg-ner-ext}

\begin{lstlisting}[language=Python, numbers=none]
system_message = """\
You are a clinical NLP assistant specializing in information extraction from medical ECG (electrocardiogram) reports. Your role is to serve as a strict, schema-aware entity extractor that produces structured annotations for downstream machine learning and clinical data analysis tasks.

Please learn the knowledge including common ECG terminologies and abbreviations first:

**Common ECG terminologies**:
Normal: "normal sinus rhythm", "normal ecg", "sinus rhythm", "within normal limits", "no abnormalities detected", ...
Abnormal: "atrial fibrillation", "ventricular tachycardia", "left ventricular hypertrophy", "right bundle branch block", "ST elevation, "T wave inversion", "prolonged QT interval", "first degree AV block", "pacemaker rhythm", ...
Uncertain: "possible infarction", "borderline ecg", "nonspecific ST-T changes", "probable left ventricular hypertrophy", "cannot rule out ischemia", ...

**Demo Abbreviations**:
NSR: "Normal Sinus Rhythm",
AFIB: "Atrial Fibrillation",
AFL: "Atrial Flutter",
VT": "Ventricular Tachycardia",
PVC: "Premature Ventricular Contraction",
PAC: "Premature Atrial Contraction",
LVH: "Left Ventricular Hypertrophy",
RVH: "Right Ventricular Hypertrophy",
RBBB: "Right Bundle Branch Block",
LBBB: "Left Bundle Branch Block",
AVB1: "First Degree AV Block",
AVB2: "Second Degree AV Block",
AVB3: "Third Degree AV Block",
STEMI: "ST-Elevation Myocardial Infarction",
NSTEMI: "Non-ST-Elevation Myocardial Infarction",
TW": "T Wave Inversion",
QTc: "Corrected QT Interval",
BBB: "Bundle Branch Block",
LAD: "Left Axis Deviation",
RAD: "Right Axis Deviation",
SA: "Sinoatrial",
PVCs: "Premature Ventricular Contractions",
PACs: "Premature Atrial Contractions"

Your primary task is to identify all relevant entities in an ECG report and then classify based on diagnosis certainty, afterwards output them in a **strictly formatted JSON object** that conforms exactly to the following schema:

```json
{
    "global": [...],    # All ECG entities from the provided report
    "classification": {
        "normal": [...],     # Entities confidently labeled as clinically "normal" (e.g., "normal ECG", "sinus rhythm")
        "abnormal": [...],   # Entities labeled as clinically "abnormal" (e.g., "atrial fibrillation", "ST elevation")
        "uncertain": [...]   # Entities with uncertainty or ambiguity in the report context (e.g., "possible LVH", "undetermined".)
    }
}
```
    
**Strict constraints**:

- Return **only** the JSON object. Do not include any natural language explanation or commentary.
- Do not hallucinate or invent fields not specified above.
- Do not extract adjectives or modifiers (e.g., "nonspecific", "mild", "marked", "possibly", "likely") as standalone entities. If a descriptive modifier qualifies an entity (e.g., "nonspecific ST-T changes", "likely normal ecg"), include it in the full entity string.
- Do not extract entire sentences or diagnostic phrases as a single entity. If a sentence contains multiple medical concepts, extract each as a separate entity.
- If an entity contains conjunctions (e.g., "and", "or", "and/or"), causal phrases (e.g., "due to", "with"), or multiple anatomical locations (e.g., "inferior/lateral"), you must split it into separate entities.
- If there are entities with clinically same meanings in the given report, only retain one with better expression.

**Some examples**:

- [Modifier + Entity]:  
  Input: "lateral st-t changes are probably due to ventricular hypertrophy"  
  Output: {"global": ["lateral st-t changes", "ventricular hypertrophy"], "classification": {"normal": [], "abnormal": ["lateral st-t changes", "ventricular hypertrophy"], "uncertain": []}}

- [Entity A with/and/or/'/' Entity B]:  
  Input: "sinus rhythm with pacs. hypertrophy and/or ischemia. inferior/lateral st-t changes."  
  Output: {"global": ["sinus rhythm", "pacs", "hypertrophy", "ischemia", "inferior st-t changes", "lateral st-t changes"], "classification": {"normal": ["sinus rhythm"], "abnormal": ["pacs", "hypertrophy", "ischemia", "inferior st-t changes", "lateral st-t changes"], "uncertain": []}}

- [Entity + Further Description]:  
  Input: "inferior infarct - age undetermined. pacemaker rhythm - no further analysis. poor r wave progression - probable normal variant."
  Output: {"global": ["inferior infarct", "age undetermined", "pacemaker rhythm", "poor r wave progression", "probable normal variant"], "classification": {"normal": [], "abnormal": ["inferior infarct", "pacemaker rhythm", "poor r wave progression"], "uncertain": ["age undetermined", "probable normal variant"]}}  # "no further analysis" is not a medical entity

Your output will be used in real-life clinical settings. Any deviation from this format may cause serious issues in downstream applications. Be precise and compliant.
"""


def get_prompt(row):
    return f"""\
Please extract all relevant clinical entities from the following ECG report.

Return the output strictly in the JSON format described in the system prompt.
Do not include any explanation or additional text.

ECG report text:
\"{row["total_report"]}\"
"""
\end{lstlisting}

\subsubsection{Case Study of Deduplication and Mapping}
\label{app:llm-ecg-ner-case}

To address concerns about how descriptive cardiac queries are constructed and how they reduce noise compared to raw NER outputs, we present a representative case study from the MIMIC-IV-ECG dataset.

\noindent\textbf{Original Clinical Report:}
\begin{quote}
“Sinus rhythm w/ PACs, QTc prolonged, Left axis deviation, RBBB with left anterior fascicular block, Inferior/lateral T changes may be due to myocardial ischemia, Low QRS voltages in precordial leads.”
\end{quote}

\noindent\textbf{Extracted Raw Entities:}
\begin{quote}
\texttt{
"sinus rhythm", "PACs", "QTc prolonged", "Left axis deviation", "RBBB", "Left anterior fascicular block", "Inferior/lateral T changes", "Myocardial ischemia", "Low QRS voltages in precordial leads"
}
\end{quote}

\noindent\textbf{Mapped and Standardized Queries (after Deduplication and Mapping):}

\begin{table}[ht]
    \centering
    \resizebox{\columnwidth}{!}{
        \begin{tabular}{l|l|l}
            \toprule[1.2pt]
            \textbf{SCP Code} & \textbf{Standardized Query} & \textbf{Matched Raw Entities (Cosine Similarity)} \\
            \midrule
            SR & sinus rhythm & sinus rhythm (1.0000) \\
            PAC & premature atrial complex & PACs (0.8976) \\
            LNGQT & prolonged QT interval & QTc prolonged (0.9434) \\
            ALS & axis left shift & Left axis deviation (0.8723) \\
            RBBB & right bundle branch block & RBBB (1.0000) \\
            LAFB & left anterior fascicular block & Left anterior fascicular block (1.0000) \\
            NT & non-specific T wave changes & Inferior/lateral T changes (0.7751) \\
            MI & myocardial infarction & Myocardial ischemia (0.9231) \\
            LVOLT & low QRS voltages & Low QRS voltages in precordial leads (0.8919) \\
            \bottomrule[1.2pt]
        \end{tabular}
    }
    \caption{Example mapping from raw NER entities to standardized cardiac query labels.}
    \label{tab:query-case}
    \vspace{-10pt}
\end{table}

This example illustrates how the same clinical concept may be expressed in different lexical forms (e.g., “PACs” vs.\ “premature atrial complex”) or contain verbose phrasing (e.g., “Low QRS voltages in precordial leads”), leading to noisy or redundant supervision if used directly. By clustering and mapping using MedCPT embeddings and similarity thresholds (Table~\ref{tab:query-case}), these expressions are unified under concise, standardized queries aligned with SCP codes.

\begin{table}[ht]
    \centering
    \resizebox{\columnwidth}{!}{
    \begin{tabular}{l|l|p{9cm}}
        \toprule[1.2pt]
        \textbf{SCP Code} & \textbf{Standard Cardiac Query} & \textbf{Mapped Raw NER Entities (Cosine Similarity)} \\
        \midrule
        NORM & normal & 
        Normal (1.000), of normal (0.995), Normal result (0.979), Normal interest (0.953), percent of normal (0.942) \\
        \midrule
        IMI & inferior myocardial infarction &
        Inferior myocardial ischemia (0.956), Inferior MI on ECG (0.935), ECG shows inferior MI (0.928), Myocardial infarction (0.923), Old inferior MI (0.907) \\
        \midrule
        LVH & left ventricle hypertrophy &
        Left ventricular hypertrophy (0.992), Severe LVH (0.967), Hypertensive LVH (0.948), Acquired LVH (0.940), Congenital LVH (0.904) \\
        \bottomrule[1.2pt]
    \end{tabular}
    }
    \caption{Example clusters of raw NER entities mapped to standardized cardiac queries.}
    \label{tab:entity_cluster_table}
    \vspace{-10pt}
\end{table}

In Table~\ref{tab:entity_cluster_table} we show parts of the clustering and deduplication results on the pre-train dataset MIMIC-IV-ECG. This process prevents redundant terms from introducing duplicate supervision, normalizes entities with modifiers (e.g., “in precordial leads”), and enforces semantic consistency across ECG samples. These standardized queries form the global label set used for training, enabling clean multimodal supervision and robust generalization in zero-shot settings.

\subsection{Pre-training Framework Implementation}
\label{app:implementation}

\subsubsection{Transformer Block Structure.}
\label{app:implementation-trm}

\begin{figure}[h]
    \begin{center}
        \includegraphics[width=0.75\columnwidth]{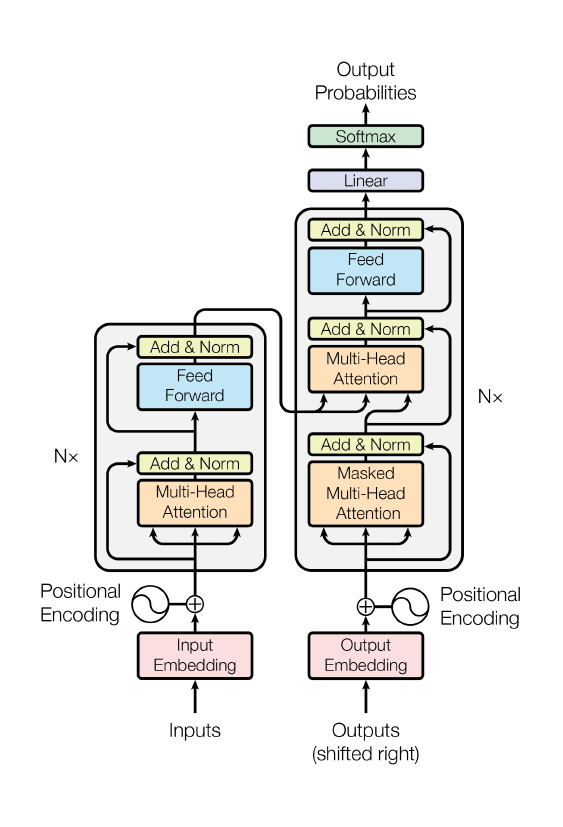}
    \end{center}
    \vspace{-15pt}
    \caption{Encoder-decoder structure of Transformer, quoted from \citep{vaswani2017attention}.}
    \label{fig:transformer}
\end{figure}

The Transformer architecture \citep{vaswani2017attention} is widely used for seq2seq modeling, learning global dependencies via self-attention instead of recurrent or convolutional structures. It consists of an encoder-decoder design, where both the encoder and decoder utilize stacked self-attention and feed-forward layers, as shown in Figure~\ref{fig:transformer}.

Each encoder block applies a residual connection around its multi-head self-attention ($\mathbf{MHA}$) and position-wise feed-forward ($\mathbf{FF}$) sublayers, followed by layer normalization:

\begin{equation}
    \begin{array}{c}
        \mathbf{Z}^{(k,1)} = \mathbf{Z}^{(k-1)} + \text{Drop}(\mathbf{MHA}(\text{Norm}(\mathbf{Z}^{(k-1)}))) \\[0.5em]
        \mathbf{Z}^{(k,2)} = \mathbf{Z}^{(k,1)} + \text{Drop}(\mathbf{FF}(\text{Norm}(\mathbf{Z}^{(k,1)}))) \\[0.5em]
        \mathbf{Z}^{\text{norm}} = \text{Norm}(\mathbf{Z}^{\text{(final)}})
    \end{array}
\end{equation}
\vspace{-10pt}

where $\mathbf{Z}^{(k-1)}$ is the input to the $k$-th Transformer block, $\mathbf{Z}^{(k,1)}$ represents the intermediate state after multi-head attention and residual connection, and $\mathbf{Z}^{(k,2)}$ is the output after the feed-forward network. The final normalized representation $\mathbf{Z}^{\text{norm}}$ is used for downstream ECG classification.

The decoder extends the encoder structure by introducing an additional multi-head attention sublayer that attends to encoder outputs, while also incorporating masked self-attention to ensure autoregressive sequence modeling. These layers collectively enable flexible cardiac feature extraction in SuPreME.

\subsubsection{Projection of ECG Embeddings}
\label{app:implementation-ecg-proj}

Following the Transformer encoder stack in the Vision Transformer (ViT) backbone, the resulting ECG token sequence $\mathbf{Z} \in \mathbb{R}^{B \times (L \cdot N) \times D}$ is passed through a modality-specific projection head to align its dimensionality with the shared multimodal latent space used in fusion.

The projection head is implemented as a two-layer multilayer perceptron ($\text{MLP}_{\text{ECG}}$), consisting of:

\begin{itemize}
    \item A linear transformation from the ViT output width $D$ to an intermediate hidden size $D_h$;
    \item A non-linear activation function (ReLU);
    \item A linear transformation from $D_h$ to the final projected dimension $D'$, shared with the text modality.
\end{itemize}

Formally, the projection can be written as:

\begin{equation}
    \begin{array}{c}
        \mathbf{Emb}_{\text{ECG}}^{\text{hidden}} = \mathbf{Z}_{\text{dropout}} \mathbf{W}_1 + \mathbf{b}_1 \\[0.5em]
        \mathbf{Emb'}_{\text{ECG}}^{\text{hidden}} = \text{ReLU}(\mathbf{Emb}_{\text{ECG}}^{\text{hidden}}) \\[0.5em]
        \mathbf{F}_{\text{ECG}} = \mathbf{Emb'}_{\text{ECG}}^{\text{hidden}} \mathbf{W}_2 + \mathbf{b}_2
    \end{array}
\end{equation}

where $\text{ReLU}$ is the activation function, $\mathbf{b}_1$ and $\mathbf{b}_2$ bias terms, and $\mathbf{W}_1 \in \mathbb{R}^{D \times D_h}$ and $\mathbf{W}_2 \in \mathbb{R}^{D_h \times D'}$ are learnable parameters.

This projection layer serves to improve non-linear representational capacity before multimodal alignment, and to map ViT-specific features to a dimensionally consistent space with text query embeddings, enabling efficient cross-modal attention in the Cardiac Fusion Network (CFN).

\subsubsection{Projection of Text Query Embeddings}
\label{app:implementation-text-proj}

To align cardiac query embeddings with ECG features in the multimodal latent space, we apply a modality-specific projection head to the output of the MedCPT query encoder (QEnc). Given $M$ queries encoded into a matrix $\mathbf{E} \in \mathbb{R}^{M \times 768}$, the projection head transforms each 768-dimensional embedding into a $D'$-dimensional representation compatible with ECG tokens.

The projection is implemented as a two-layer multilayer perceptron ($\text{MLP}_{\text{Query}}$) as well, consisting of:

\begin{itemize}
    \item A linear transformation from 768 to a hidden dimension $D_h$;
    \item A non-linear activation function (GELU);
    \item A linear transformation from $D_h$ to the target fusion dimension $D'$.
\end{itemize}

Formally, the operation is defined as:

\begin{equation}
    \begin{array}{c}
        \mathbf{Emb}_{\text{Query}}^{\text{hidden}} = \mathbf{E}_{\text{dropout}} \mathbf{W}_3 + \mathbf{b}_3 \\[0.5em]
        \mathbf{Emb'}_{\text{Query}}^{\text{hidden}} = \text{GeLU}(\mathbf{Emb}_{\text{Query}}^{\text{hidden}}) \\[0.5em]
        \mathbf{F}_{\text{Query}} = \mathbf{Emb'}_{\text{Query}}^{\text{hidden}} \mathbf{W}_4 + \mathbf{b}_4
    \end{array}
\end{equation}

where $\text{GeLU}$ is the activation function, $\mathbf{b}_3$ and $\mathbf{b}_4$ bias terms, and $\mathbf{W}_3 \in \mathbb{R}^{768 \times D_h}$ and $\mathbf{W}_4 \in \mathbb{R}^{D_h \times D'}$ are learnable parameters.

This projection head enables cross-modal alignment by transforming domain-specific textual semantics into a shared feature space used by the Cardiac Fusion Network (CFN). The structure mirrors the ECG-side projection to maintain architectural symmetry and training stability.

\subsubsection{Initialization of Cardiac Fusion Network.}
\label{app:implementation-cfn-init}

All weights in linear layers and attention modules are initialized with a normal distribution, \( \mathbf{W} \sim \mathcal{N}(0, 0.02) \). To support batch processing, the text embeddings $\mathbf{F}_\text{text}$ are expanded to match the batch size $B$. Both ECG and text embeddings undergo layer normalization to improve training stability and convergence.

\subsection{Zero-shot Evaluation Analysis}
\label{app:zeroshot-eval}

\subsubsection{Specific Classification Mechanism}

During zero-shot evaluation, the class set (i.e., diagnostic query set $\mathcal{Q}$) is dynamically specified per downstream dataset but remains fixed for all samples within that dataset. The model computes one score per query in $\mathcal{Q}$ for a given ECG sample. These scores are produced via a sigmoid-activated MLP head following the Cardiac Fusion Network (CFN) output, where each query representation attends over the ECG feature sequence. Importantly, this design supports variable-sized query sets across datasets, and prediction is always performed over the currently defined $\mathcal{Q}$. The classifier weights are not pre-defined or fixed, but learned representations aligned to query embeddings through cross-modal attention, ensuring full flexibility across unseen classes.

\subsubsection{Simplified Clinical Knowledge-Enhanced Prompt Engineering}

In our implementation of simplified CKEPE query construction, we follow the general design principle introduced in MERL~\cite{liu2024zero}. The original CKEPE pipeline in MERL employs GPT-4 with web browsing to retrieve attributes and subtypes of each cardiac condition from clinical knowledge sources such as SNOMED CT and SCP-ECG. The prompt typically used is:

\begin{quote}
\texttt{"Which attributes and subtypes does <cardiac condition> have?"}
\end{quote}

The responses are then validated against the external databases to avoid hallucination and finally organized into detailed clinical descriptions used as prompts for downstream evaluation (see MERL Section 3.4 and Figure 3).

In contrast, we adopt a simplified version of this process (Section~\ref{sec:zero-shot}) aimed at reducing verbosity while preserving clinical specificity. Specifically, we use GPT-4o with the following style of prompt:

\begin{quote}
\texttt{"Provide the standard clinical definition of <SCP diagnostic code> based on the SCP-ECG protocol."}
\end{quote}

The generated responses are then automatically validated by external databases as well to reduce hallucinated content. Rather than expanding into all potential attributes or phenotypes (as done in MERL), we retain only the concise, high-precision diagnostic phrase for each class, enabling cleaner alignment with the downstream label space.

Take a simple case study as example, for the diagnostic class LBBB (Left Bundle Branch Block), MERL would produce a long-form prompt such as:

\begin{quote}
“A conduction abnormality characterized by delayed depolarization of the left ventricle, typically resulting in a widened QRS complex ($>$120 ms), often associated with underlying structural heart disease or ischemia.”
\end{quote}

In contrast, our simplified prompt (after GPT-4o generation and medical verification) becomes:

\begin{quote}
“left bundle branch block”
\end{quote}

This compact form reduces potential noise in query encoding while retaining diagnostic specificity. It aligns with our hypothesis that multimodal fusion benefits more from semantically discriminative labels than verbose natural language definitions.

\subsubsection{Evaluation Metrics}

We use zero-shot learning and linear probing to evaluate the performance of our framework and mainstream eSSL frameworks. The primary metric is Area Under the Receiver Operating Characteristic (AUROC, also referred to as AUC). AUROC is widely used to evaluate the performance of binary classification models. The ROC curve plots the True Positive Rate (TPR) on the vertical axis against the False Positive Rate (FPR) on the horizontal axis. By varying the classifier's threshold, $\mathbf{TPR}$ and $\mathbf{FPR}$ are calculated and then plotted to form the curve, where $\mathbf{TP}$ refers to True Positive, $\mathbf{FN}$ refers to False Negative, $\mathbf{FP}$ refers to False Positive, and $\mathbf{TN}$ refers to True Negative.:

\begin{equation}
    \begin{array}{c}
        \mathbf{TPR} = \dfrac{\mathbf{TP}}{\mathbf{TP} + \mathbf{FN}} \\ \\
        \mathbf{FPR} = \dfrac{\mathbf{FP}}{\mathbf{FP} + \mathbf{TN}}
    \end{array}
\end{equation}

AUROC is the area under the ROC curve, with values ranging from 0 to 1, reflecting the overall classification ability of the model. \textbf{AUROC = 0.5} indicates that the model's classification ability is equivalent to random guessing, while \textbf{AUROC $>$ 0.5} and values closer to \textbf{1} indicate that the model is able to classify with greater accuracy.

\subsection{Statistics and Overlap Analysis}

\subsubsection{Statistics of Extracted MIMIC-IV-ECG Entities}
\label{app:mimic-stats}

We extract over 3.4 million clinical entities from free-text ECG reports in the MIMIC-IV-ECG dataset using an instruction-tuned LLM. At the term level, this results in 1,168 unique raw entities (Table~\ref{tab:entity_stats}). Among these, 93.75\% remain after filtering out uncertain or ambiguous expressions. To resolve redundancy and lexical variation, we apply embedding-based clustering using MedCPT representations, reducing the vocabulary to 341 cluster representatives. Further manual verification and mapping to UMLS/SNOMED CT terminologies yield a final set of 295 standardized cardiac entities used as global queries during supervised pre-training.

\begin{table}[ht]
    \centering
        \resizebox{\columnwidth}{!}{
            \begin{tabular}{l|c|c}
                \toprule[1.2pt]
                \textbf{Entity Type} & \textbf{Count} & \textbf{Proportion} \\
                \midrule
                Raw extracted entities & 3,419,064 & 100\% (sample-level) \\
                Unique raw extracted entities & 1,168 & 100\% (term-level) \\
                Terms after uncertainty filtering & 1,095 & 93.75\% (vs. 1,168) \\
                Entity cluster representatives (post-deduplication) & 341 & 29.20\% (vs. 1,168) \\
                Final unique standardized entities (post-mapping) & 295 & 25.26\% (vs. 1,168) \\
                \bottomrule[1.2pt]
            \end{tabular}
        }
    \caption{Statistics of unique cardiac Entities: Extraction, Filtering, Deduplication, and Mapping}
    \label{tab:entity_stats}
    \vspace{-10pt}
\end{table}

\begin{figure}[h]
    \centering
    \includegraphics[width=\columnwidth]{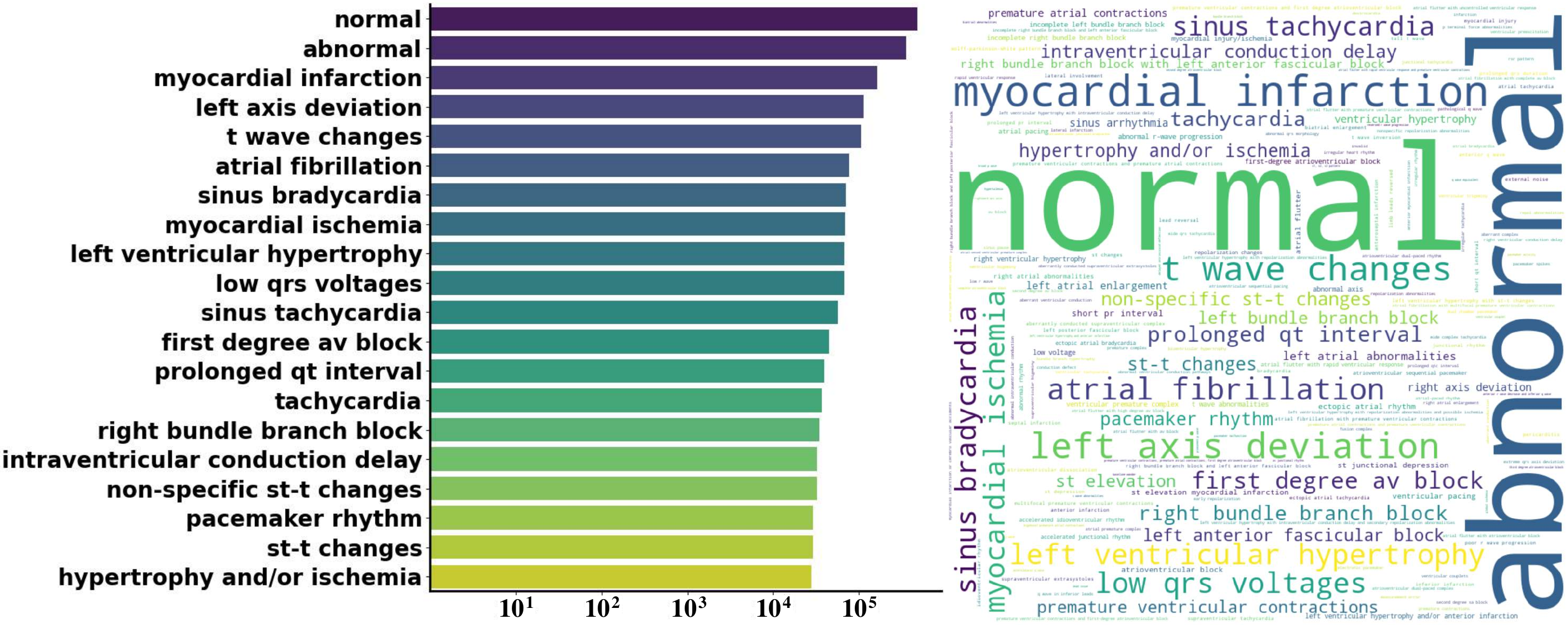}
    \vskip -0.1in
    \caption{Frequency distribution of standardized ECG entities after deduplication and mapping in MIMIC-IV-ECG.}
    \label{fig:freq}
    \vspace{-5pt}
\end{figure}

Table~\ref{tab:cluster_stats} provides additional statistics on the clustering process. The average cluster contains 3.39 entities, with some clusters merging up to 29 semantically similar terms. In total, 86.51\% of clusters are successfully mapped to standardized terms. The distribution of standardized entity frequencies is illustrated in Figure~\ref{fig:freq}. The left panel shows a log-scaled histogram of the most common cardiac terms, with "normal", "abnormal", and "myocardial infarction" being the most frequent. The right panel presents a word cloud that qualitatively reflects term prevalence and semantic variety. Together, these visualizations confirm that while a few diagnostic terms dominate the corpus, a long tail of clinically significant but less frequent entities is preserved, supporting robust coverage in downstream classification.

\begin{table}[ht]
    \centering
        \resizebox{\columnwidth}{!}{
            \begin{tabular}{l|c}
                \toprule[1.2pt]
                \textbf{Clustering Metric} & \textbf{Value} \\
                \midrule
                Number of clusters formed & 341 \\
                Average number of entities per cluster & 3.39 \\
                Maximum / Minimum cluster size & 29 / 1 \\
                Proportion of clusters mapped to standard terms & 86.51\% \\
                \bottomrule[1.2pt]
            \end{tabular}
        }
    \caption{Clustering statistics of extracted cardiac entities on MedCPT embeddings}
    \label{tab:cluster_stats}
\end{table}

\subsubsection{Dataset Overlap Analysis}
\label{app:overlap}

\begin{figure}[h]
    \includegraphics[width=0.48\linewidth]{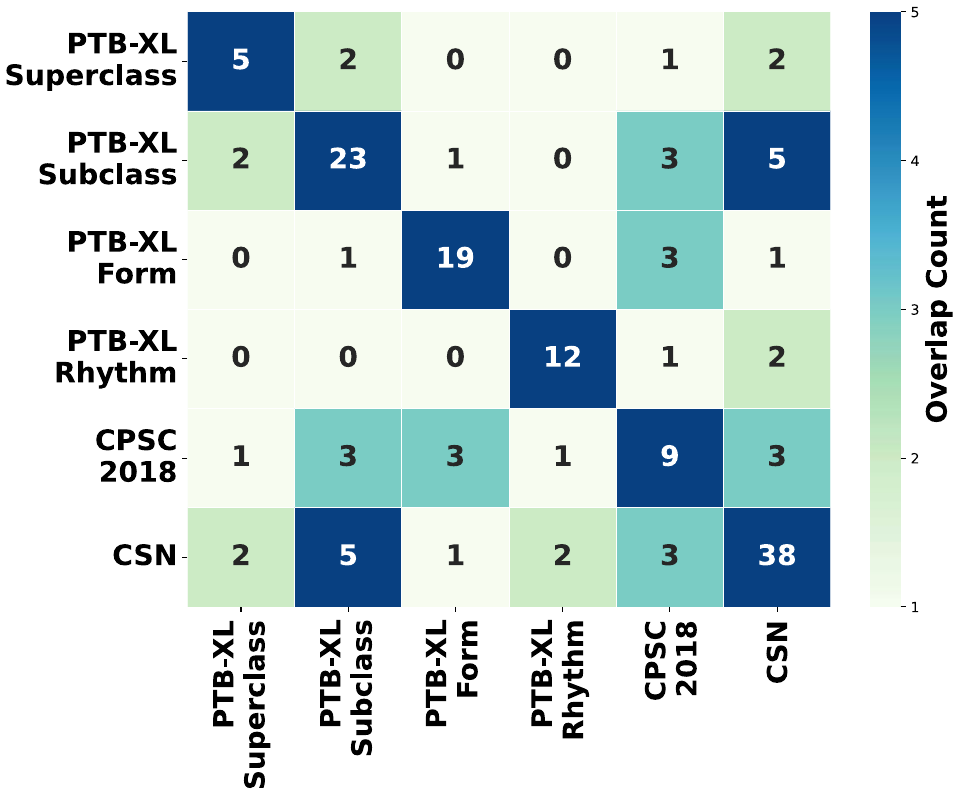} \hfill
    \includegraphics[width=0.48\linewidth]{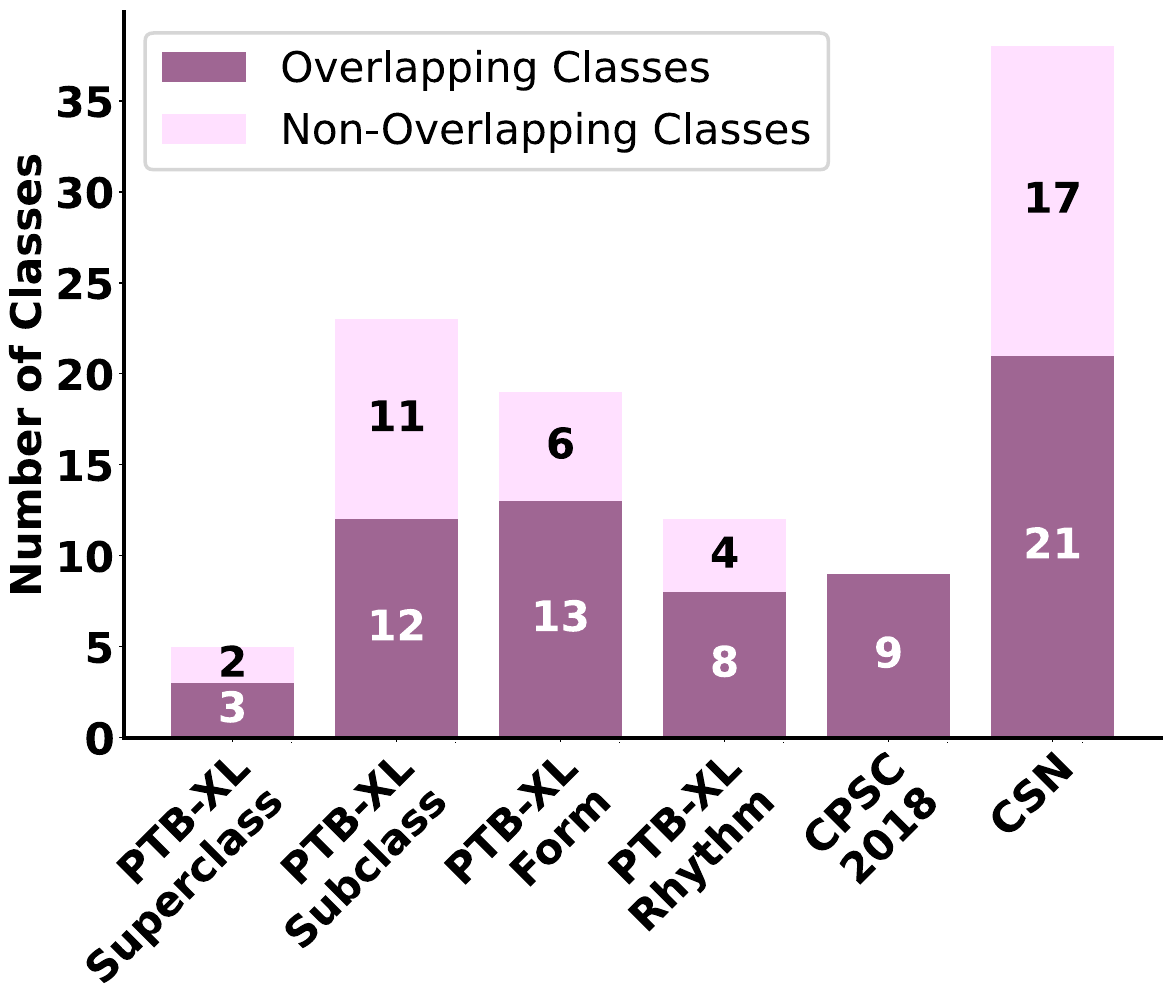}
    \caption {Overlap between ECG datasets, with left panel showing pairwise overlap counts between downstream datasets, and right panel showing the distribution of overlapping and non-overlapping classes between each downstream dataset and the pre-training dataset.}
    \label{fig:overlap}
\end{figure}

We analyze the cardiac query overlap between the pre-training dataset and six downstream datasets specified in Section~\ref{sec:config4exp}, as well as among the downstream datasets themselves, as illustrated in Figure~\ref{fig:overlap}. Specifically, we embed all entities from the pre-training dataset and cardiac queries from the downstream datasets, compute their cosine similarity, and apply a threshold of 0.95 verified by cardiologists with 10+ years of experience as well to filter overlapping queries. 

The heatmap on the left shows that pairwise overlaps among downstream datasets are generally limited, reflecting the diversity of cardiac query prompts. The bar chart on the right reveals that 57 cardiac queries overlap between the pre-training dataset and the downstream datasets. Despite the pre-training dataset shares some similar queries, a substantial portion of queries remains unique to the downstream datasets, allowing the pre-training process to establish robust general-purpose representations while leaving room for downstream-specific adaptation.

Table~\ref{tab:overlap} shows the overlap between entities from the pre-training dataset and cardiac queries from the downstream datasets, filtered using a cosine similarity threshold of 0.95.

\begin{table}[h]
    \centering
    \resizebox{\columnwidth}{!}{
        \begin{tabular}{l|l|c}
            \toprule[1.2pt]
            \textbf{Pre-training Dataset Entity}      & \textbf{Downstream Cardiac Query}    & \textbf{Similarity Score} \\ 
            \midrule[1.2pt]
            atrial fibrillation                    & atrial fibrillation                  & 1.0000                    \\ 
            supraventricular tachycardia           & supraventricular tachycardia         & 1.0000                    \\ 
            ventricular preexcitation              & ventricular preexcitation            & 1.0000                    \\ 
            right bundle branch block              & right bundle branch block            & 1.0000                    \\ 
            myocardial infarction                  & myocardial infarction                & 1.0000                    \\ 
            atrial premature complex               & atrial premature complex             & 1.0000                    \\ 
            Prolonged QT interval                  & prolonged qt interval                & 1.0000                    \\ 
            T wave abnormalities                   & t wave abnormalities                 & 1.0000                    \\ 
            ST depression                          & st depression                        & 1.0000                    \\ 
            AV block                               & av block                             & 1.0000                    \\ 
            T wave Changes                         & t wave changes                       & 1.0000                    \\ 
            sinus bradycardia                      & sinus bradycardia                    & 1.0000                    \\ 
            left anterior fascicular block         & left anterior fascicular block       & 1.0000                    \\ 
            sinus arrhythmia                       & sinus arrhythmia                     & 1.0000                    \\ 
            left bundle branch block               & left bundle branch block             & 1.0000                    \\ 
            sinus tachycardia                      & sinus tachycardia                    & 1.0000                    \\ 
            abnormal Q wave                        & abnormal q wave                      & 1.0000                    \\ 
            ventricular premature complex          & ventricular premature complex        & 1.0000                    \\ 
            Prolonged PR interval                  & prolonged pr interval                & 1.0000                    \\ 
            Atrial Tachycardia                     & atrial tachycardia                   & 1.0000                    \\ 
            Supraventricular Tachycardia           & supraventricular tachycardia         & 1.0000                    \\ 
            left posterior fascicular block        & left posterior fascicular block      & 1.0000                    \\ 
            normal                                 & normal                               & 1.0000                    \\ 
            second degree AV block                 & second degree av block               & 1.0000                    \\ 
            anterior myocardial infarction         & anterior myocardial infarction       & 1.0000                    \\ 
            incomplete left bundle branch block    & incomplete left bundle branch block  & 1.0000                    \\ 
            incomplete right bundle branch block   & incomplete right bundle branch block & 1.0000                    \\ 
            ST elevation                           & st elevation                         & 1.0000                    \\ 
            atrial flutter                         & atrial flutter                       & 1.0000                    \\ 
            Sinus Tachycardia                      & sinus tachycardia                    & 1.0000                    \\ 
            Sinus Bradycardia                      & sinus bradycardia                    & 1.0000                    \\ 
            first degree AV block                  & first degree av block                & 1.0000                    \\ 
            premature complex                      & premature complex                    & 1.0000                    \\ 
            ST-T change                            & st-t changes                         & 0.9968                    \\ 
            premature atrial complex               & atrial premature complex             & 0.9961                    \\ 
            left ventricle hypertrophy             & left ventricular hypertrophy         & 0.9924                    \\ 
            right ventricle hypertrophy            & right ventricular hypertrophy        & 0.9920                    \\ 
            Q wave present                         & q wave                               & 0.9903                    \\ 
            complete right bundle branch block     & right bundle branch block            & 0.9891                    \\ 
            high QRS voltage                       & high qrs voltages                    & 0.9878                    \\ 
            complete left bundle branch block      & left bundle branch block             & 0.9861                    \\ 
            second degree AV block(Type one)       & second degree av block               & 0.9817                    \\ 
            anteroseptal myocardial infarction     & anteroseptal infarction              & 0.9809                    \\ 
            ischemic                               & ischemia                             & 0.9804                    \\ 
            second degree AV block(Type two)       & second degree av block               & 0.9795                    \\ 
            third degree av block                  & second degree av block               & 0.9795                    \\ 
            low amplitude T wave                   & high t wave amplitude                & 0.9741                    \\ 
            abnormal QRS                           & abnormal qrs morphology              & 0.9737                    \\ 
            suggests digitalis-effect              & digitalis effect                     & 0.9726                    \\ 
            supraventricular arrhythmia            & supraventricular tachycardia         & 0.9684                    \\ 
            anterolateral myocardial infarction    & anterolateral infarction             & 0.9667                    \\ 
            paroxysmal supraventricular tachycardia & supraventricular tachycardia         & 0.9611                    \\ 
            left front bundle branch block         & left bundle branch block             & 0.9537                    \\ 
            inferior myocardial infarction         & inferior infarction                  & 0.9512                    \\ 
            right atrial hypertrophy               & right atrial enlargement             & 0.9570                    \\ 
            \bottomrule[1.2pt]
        \end{tabular}
    }
\caption{Overlap between pre-training dataset entities and downstream cardiac queries, filtered with similarity threshold of 0.95, sorted by similarity score.}
\label{tab:overlap}
\vspace{-20pt}
\end{table}

\subsection{Downstream Task Configuration}
\label{app:eval}

\subsubsection{Downstream Data Split}
\label{app:datasplit}

For PTB-XL, we adopt the official train-test split recommended by the dataset authors \citep{wagner2020ptb}, ensuring consistency with prior works and a balanced distribution of ECG categories. This split is directly applied across the Superclass, Subclass, Form, and Rhythm subsets of PTB-XL. For CPSC-2018 and CSN, we follow the data splitting approach used by \citep{liu2024zero}, which randomly divides the datasets into training, validation, and testing subsets in a 70\%:10\%:20\% ratio.

Details of the splits, including the specific number of samples allocated to each subset, are summarized in Table \ref{tab:datasplit}.

\begin{table}[ht]
\centering
\resizebox{\columnwidth}{!}{
    \begin{tabular}{lcccc}
        \toprule[1.2pt]
        Dataset & Category Number & Train Set & Validation Set & Test Set \\ 
        \midrule[1.2pt]
        \textbf{\textit{PTB-XL}} \\
        \midrule
        Superclass & 5 & 17,084 & 2,146 & 2,158 \\
        Subclass & 23 & 17,084 & 2,146 & 2,158 \\
        Form & 19 & 7,197 & 901 & 880 \\
        Rhythm & 12 & 16,832 & 2,100 & 2,098 \\
        \midrule
        \textbf{\textit{Others}} \\
        \midrule
        CPSC-2018 & 9 & 4,950 & 551 & 1,376 \\
        CSN & 38 & 16,546 & 1,860 & 4,620 \\
         \bottomrule[1.2pt]
    \end{tabular}
}
\caption{Data splits and sample distribution for downstream datasets.}
\label{tab:datasplit}
\vspace{-20pt}
\end{table}

\subsubsection{Downstream Experiment Configuration}
\label{app:downstream}

The training configurations for downstream tasks, including optimizer, scheduler, and relevant hyperparameters, are detailed in Table \ref{tab:hyper}.

\begin{table*}[tb!]
    \centering
    \resizebox{\textwidth}{!}{
        \begin{tabular}{lcccccc}
            \toprule[1.2pt]
            Hyperparameter & PTB-XL-Superclass & PTB-XL-Subclass & PTB-XL-Form & PTB-XL-Rhythm & CPSC-2018 & CSN\\
            \midrule[1.2pt]
            \textbf{\textit{Optimizer}} \\
            \midrule
            Type & AdamW & AdamW & AdamW & AdamW & AdamW & AdamW \\
            Learning Rate & 1e-3 & 1e-3 & 1e-3 & 1e-3 & 1e-3 & 1e-3 \\
            Weight Decay & 1e-8 & 1e-8 & 1e-8 & 1e-8 & 1e-8 & 1e-8 \\
            \midrule
            \textbf{\textit{Scheduler}} \\
            \midrule
            Type & Cosine Annealing & Cosine Annealing & Cosine Annealing & Cosine Annealing & Cosine Annealing & Cosine Annealing \\
            Warmup Steps & 5 & 5 & 5 & 5 & 5 & 5 \\
            \midrule
            \textbf{\textit{General}} \\
            \midrule
            Batch Size & 16 & 16 & 16 & 16 & 16 & 16 \\
            Epochs & 100 & 100 & 100 & 100 & 100 & 100 \\
            \bottomrule[1.2pt]
        \end{tabular}
    }
\caption{Downstream dataset information and split proportions.}
\label{tab:hyper}
\end{table*}

\subsection{Performance of Non-overlapped Cardiac Conditions}
\label{app:non-overlap}

While evaluating performance exclusively on non-overlapping (i.e., unseen) downstream classes is not a standard practice in existing ECG literature, including MERL and other multimodal or self-supervised frameworks, we acknowledge its value in assessing true generalization. To address this, we conduct an additional analysis where we evaluate zero-shot AUC only on downstream classes that do not appear in the pre-training dataset.

Table~\ref{tab:nonoverlap-eval} presents the comparison between AUC scores on all downstream classes versus only non-overlapping ones. As expected, performance on unseen classes is moderately lower, yet remains strong across datasets, confirming our framework’s ability to generalize beyond pre-trained diagnostic categories. This analysis complements our main results and provides deeper insights into model robustness.

\begin{table*}[ht]
    \centering
    \resizebox{\linewidth}{!}{
        \begin{tabular}{l|cccccc|c}
            \toprule[1.2pt]
            \textbf{Setting} & \textbf{PTB-XL-Super} & \textbf{PTB-XL-Sub} & \textbf{PTB-XL-Form} & \textbf{PTB-XL-Rhythm} & \textbf{CPSC-2018} & \textbf{CSN} & \textbf{Overall} \\
            \midrule
            Non-overlapping Classes & 75.97 & 69.30 & 61.36 & 83.83 & -- & 75.73 & 73.24 \\
            All Classes & 78.20 & 77.52 & 60.67 & 86.79 & 79.83 & 80.17 & 77.20 \\
            \bottomrule[1.2pt]
        \end{tabular}
    }
    \caption{Zero-shot AUC on downstream datasets using only non-overlapping (unseen) classes vs.\ using all classes.}
    \label{tab:nonoverlap-eval}
\end{table*}

\subsection{Performance on Specific Cardiac Conditions}
\label{app:scatters}

\begin{figure*}[h]
    \includegraphics[width=\linewidth]{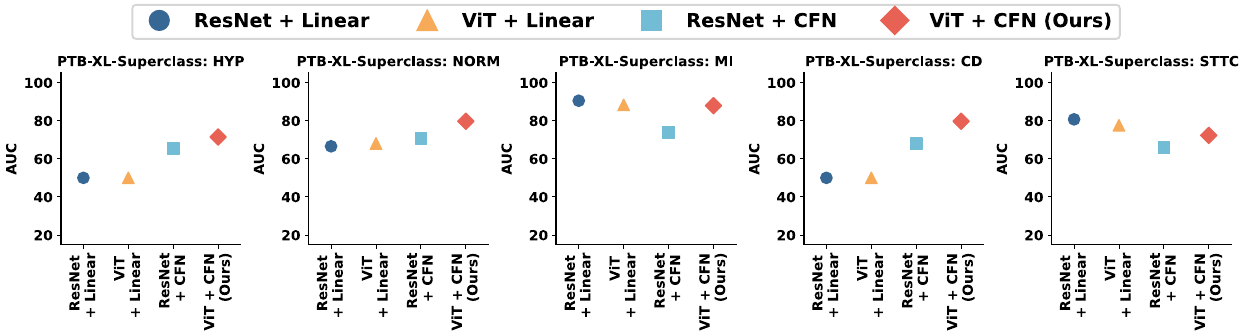}
    \vspace{-15pt}
    \caption{Zero-shot learning performance of SuPreME and its variants on specific categories in PTB-XL-Superclass.}
    \label{fig:scatter_ptbxl_superclass}
\end{figure*}

\noindent \textbf{PTB-XL-Superclass.} Figure~\ref{fig:scatter_ptbxl_superclass} records the AUC performance of SuPreME on specific cardiac conditions in PTB-XL-Superclass dataset.

\noindent \textbf{PTB-XL-Subclass.} Figure~\ref{fig:scatter_ptbxl_subclass} records the AUC performance of SuPreME on specific cardiac conditions in PTB-XL-Subclass dataset.

\noindent \textbf{PTB-XL-Form.} Figure~\ref{fig:scatter_ptbxl_form} records the AUC performance of SuPreME on specific cardiac conditions in PTB-XL-Form dataset.

\noindent \textbf{PTB-XL-Rhythm.} Figure~\ref{fig:scatter_ptbxl_rhythm} records the AUC performance of SuPreME on specific cardiac conditions in PTB-XL-Rhythm dataset.

\noindent \textbf{CPSC-2018.} Figure~\ref{fig:scatter_cpsc} records the AUC performance of SuPreME on specific cardiac conditions in CPSC-2018 dataset.

\noindent \textbf{CSN.} Figure~\ref{fig:scatter_csn} records the AUC performance of SuPreME on specific cardiac conditions in CSN dataset.

\begin{figure*}[h]
    \includegraphics[width=\linewidth]{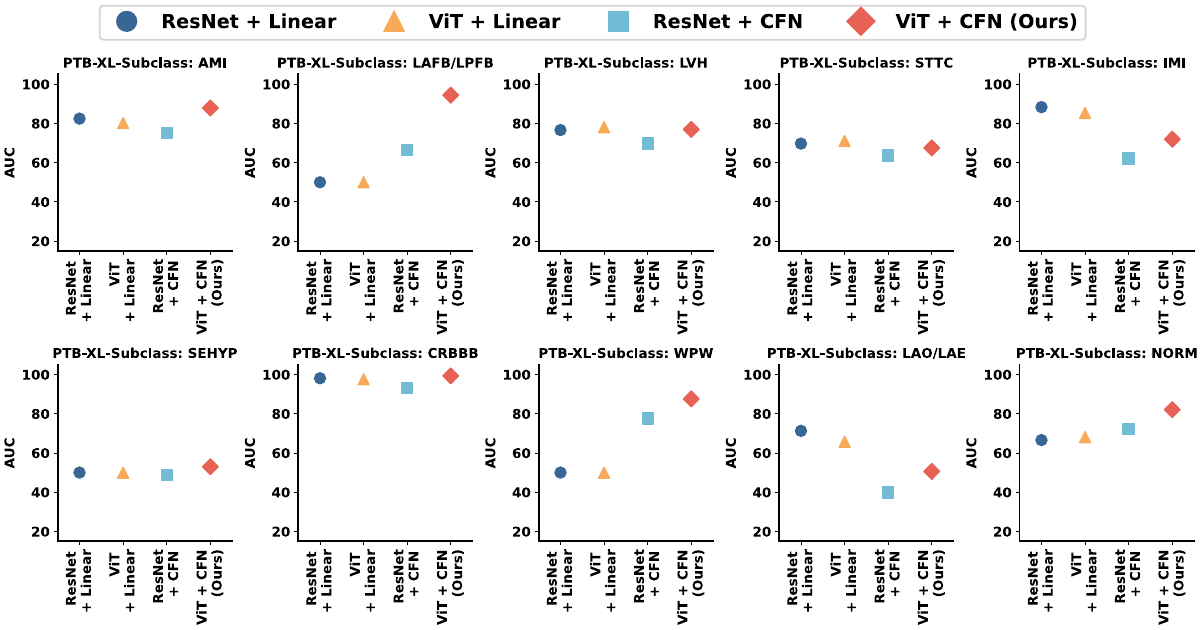}
\end{figure*}

\clearpage

\begin{figure*}[h]
    \includegraphics[width=\linewidth]{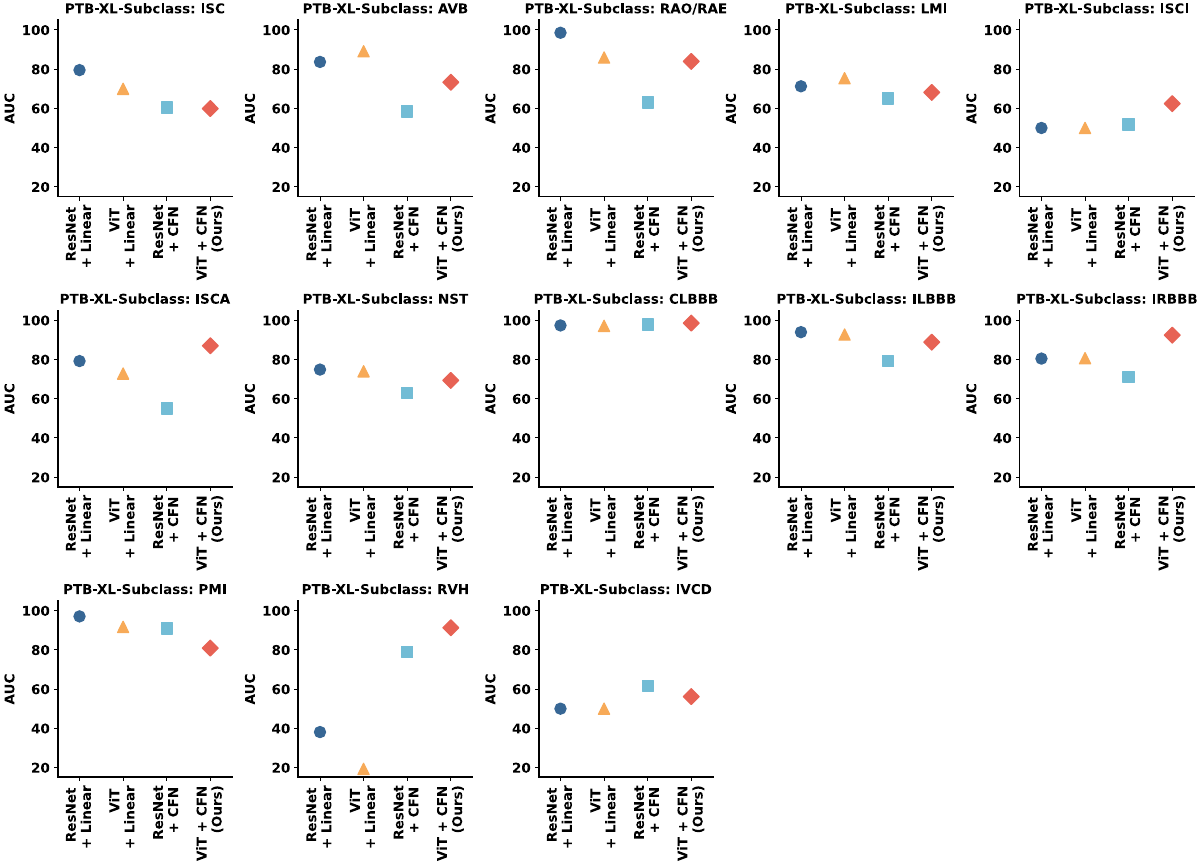}
    \vspace{-15pt}
    \caption{Zero-shot learning performance of SuPreME and its variants on specific categories in PTB-XL-Subclass.}
    \label{fig:scatter_ptbxl_subclass}
\end{figure*}

\begin{figure*}[h]
    \includegraphics[width=\linewidth]{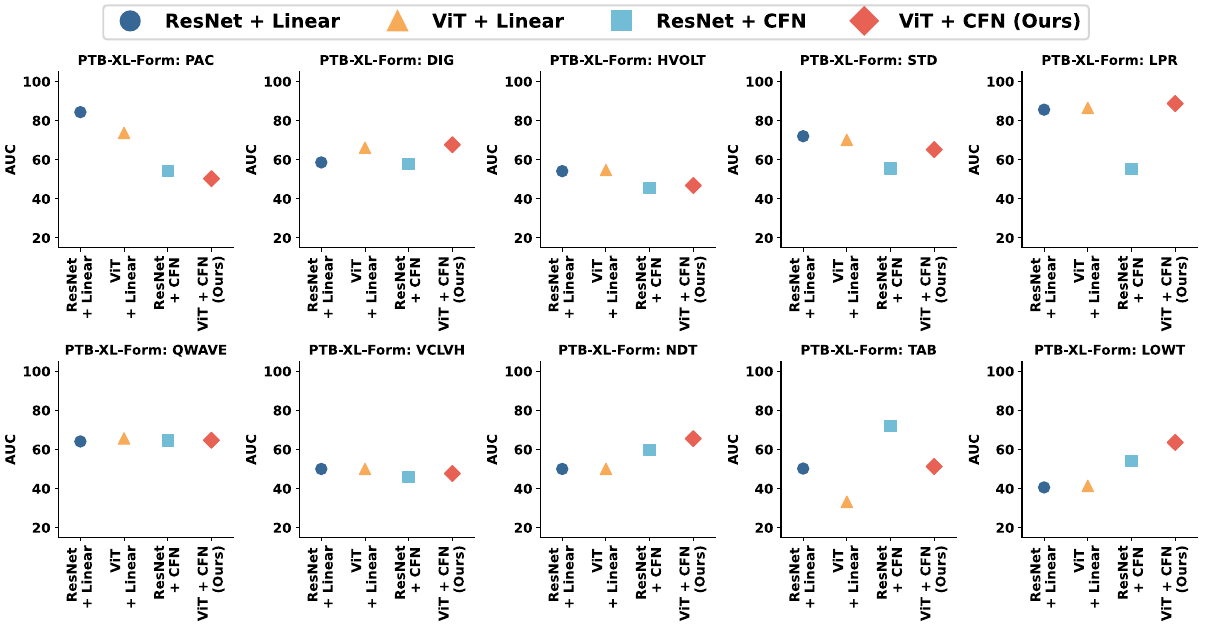}
\end{figure*}

\clearpage

\begin{figure*}[h]
    \includegraphics[width=\linewidth]{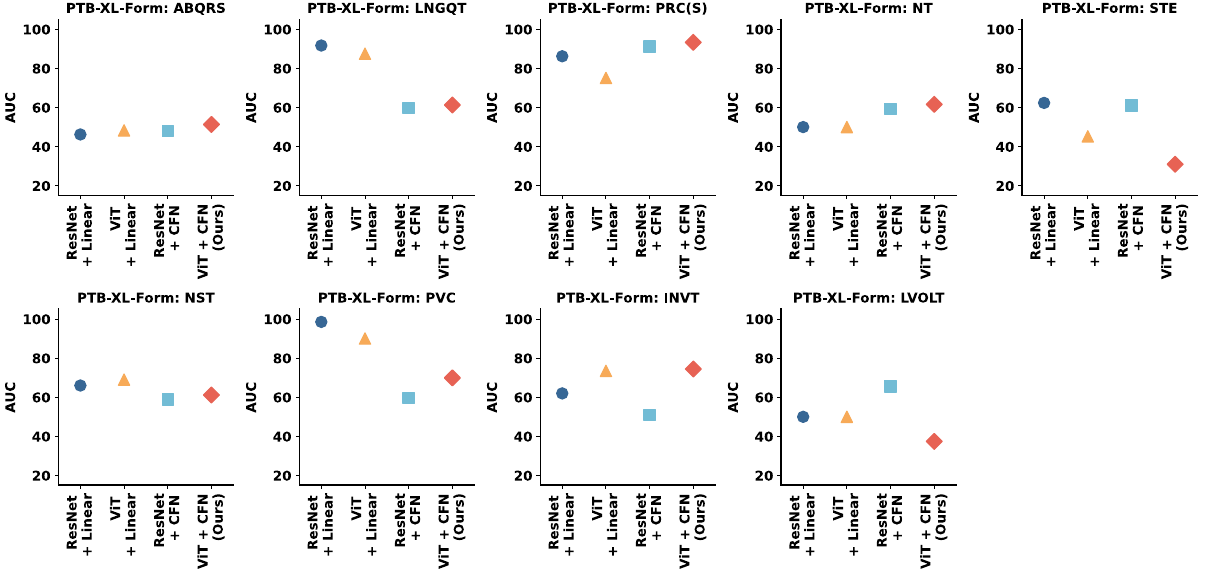}
    \vspace{-15pt}
    \caption{Zero-shot learning performance of SuPreME and its variants on specific categories in PTB-XL-Form.}
    \label{fig:scatter_ptbxl_form}
\end{figure*}

\begin{figure*}[h]
    \includegraphics[width=\linewidth]{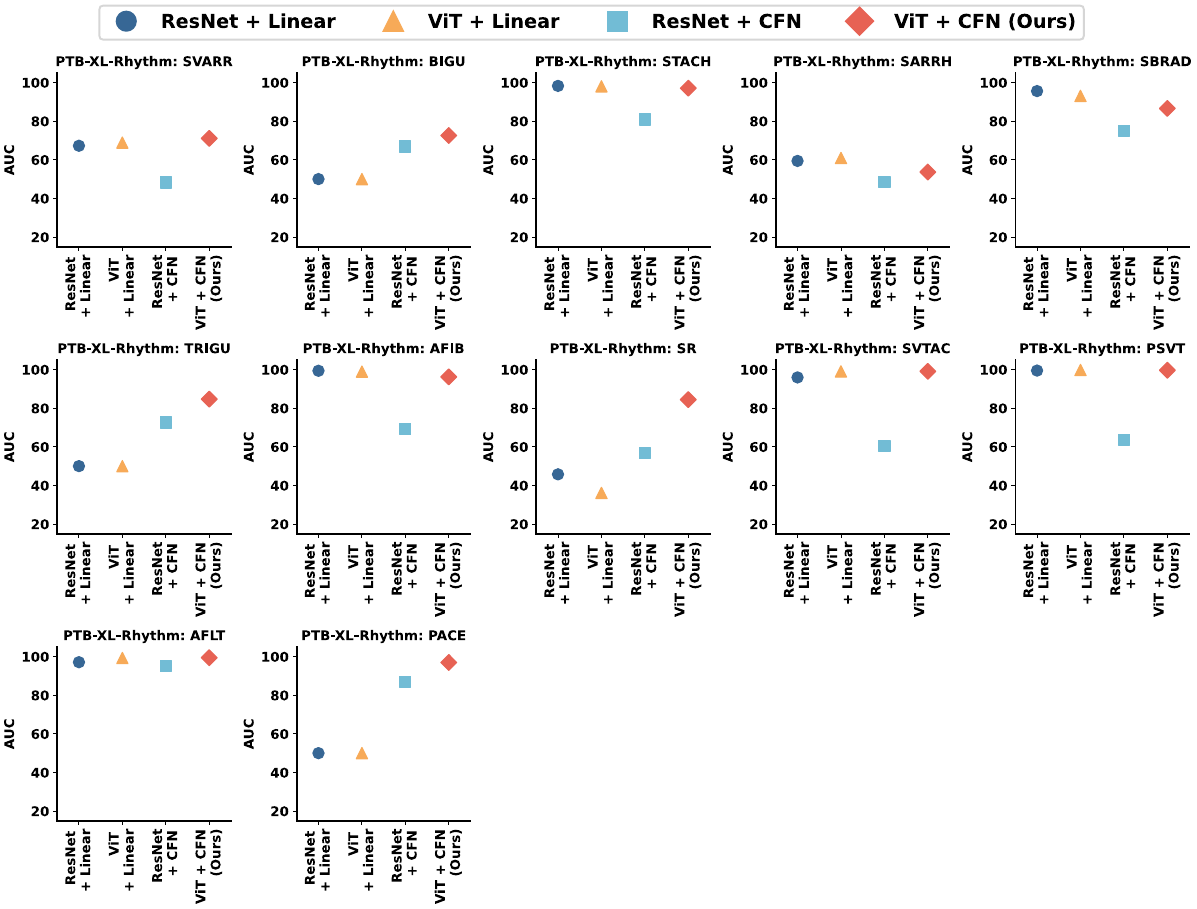}
    \vspace{-15pt}
    \caption{Zero-shot learning performance of SuPreME and its variants on specific categories in PTB-XL-Rhythm.}
    \label{fig:scatter_ptbxl_rhythm}
\end{figure*}

\clearpage

\begin{figure*}[h]
    \includegraphics[width=\linewidth]{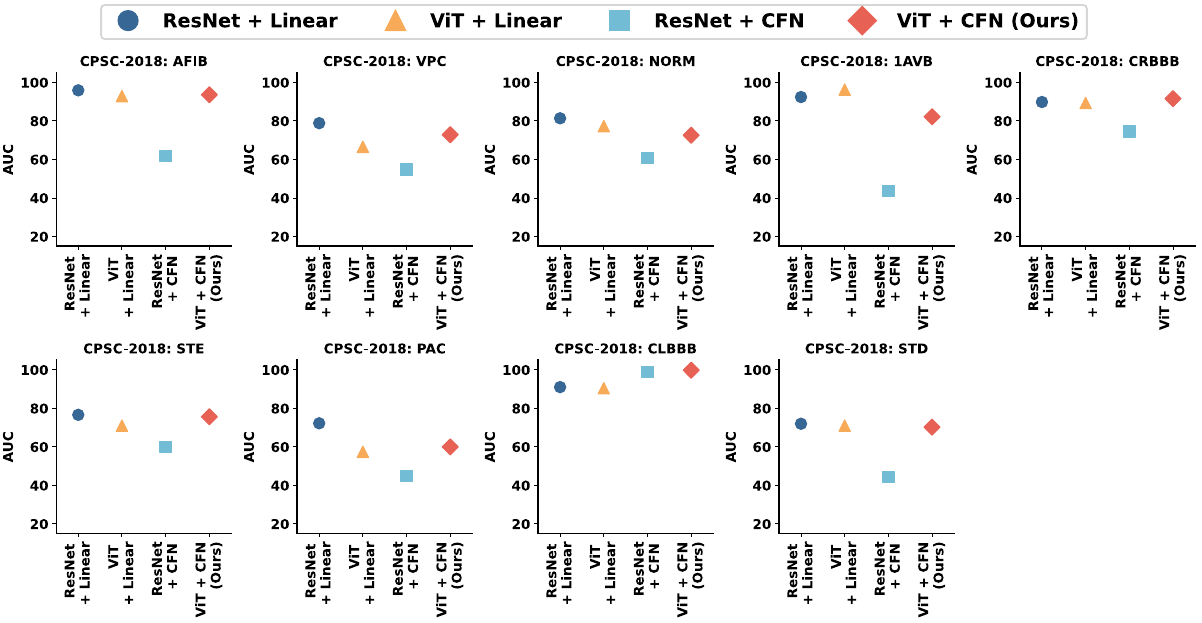}
    \vspace{-15pt}
    \caption{Zero-shot learning performance of SuPreME and its variants on specific categories in CPSC-2018.}
    \label{fig:scatter_cpsc}
\end{figure*}

\begin{figure*}[h]
    \includegraphics[width=\linewidth]{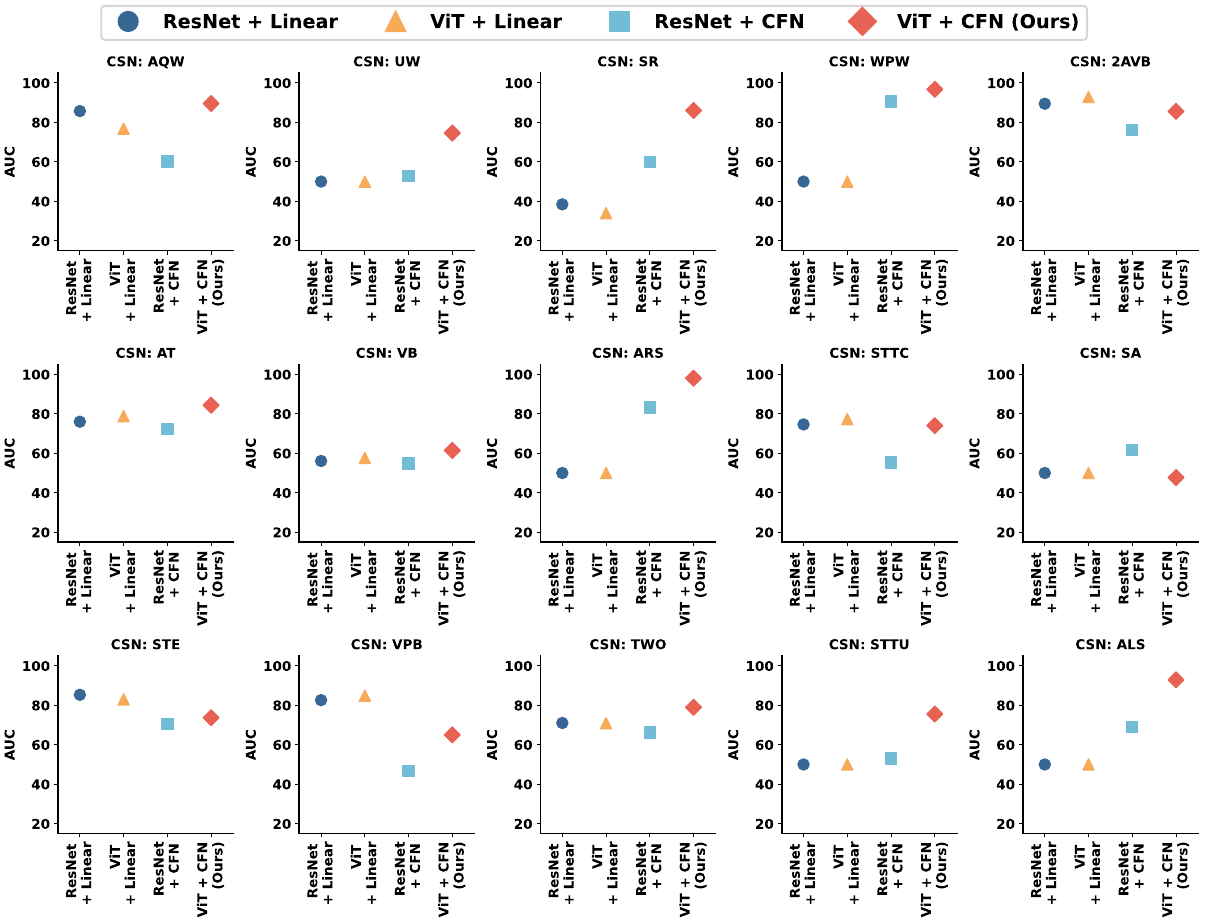}
\end{figure*}

\clearpage

\begin{figure*}[htb!]
    \includegraphics[width=\linewidth]{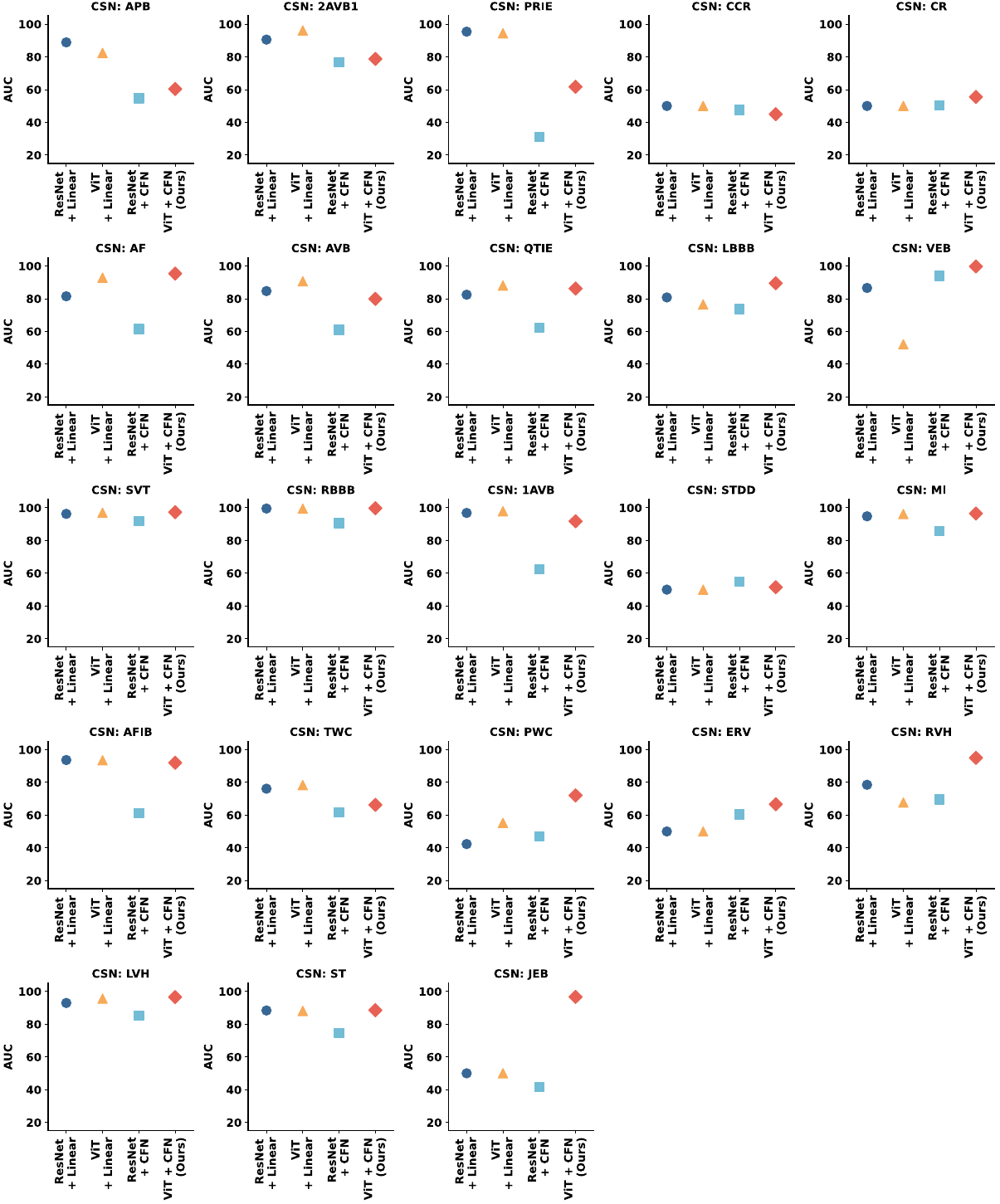}
    \vspace{-15pt}
    \caption{Zero-shot learning performance of SuPreME and its variants on specific categories in CSN.}
    \label{fig:scatter_csn}
\end{figure*}

\clearpage

\subsection{Further Discussion}

\subsubsection{Offline Use of Large-sized LLM for Clinical NER}
\label{app:dis-llm}

While we employ LLaMA3.1-70B-Instruct for clinical entity extraction, this step is performed offline only once during dataset construction and is not part of the SuPreME model's pre-training \& inference pipeline. The motivation for using a larger model is to ensure high annotation quality for the pre-training dataset. Once entities are extracted and mapped, they form a standardized query list used throughout training and evaluation. Therefore, clinical deployments do not require access to large LLMs, and the SuPreME model itself remains lightweight during inference.

\subsubsection{Computation Cost and Practical Deployment}
\label{app:dis-compute}

\noindent \textbf{Data Processing (Offline NER).}
To obtain high-quality supervision labels, we extract and normalize diagnostic entities from MIMIC-IV-ECG reports using LLaMA3.1–70B-Instruct with structured prompts. This step is performed only once as described above before pre-training SuPreME model. The output is a cleaned, deduplicated dataset of standardized diagnostic labels, which serves as supervision for training. The annotation process takes approximately 6 hours on 8 NVIDIA A100-SMX4-80GB GPUs, and the tested minimum reproducing resources are 4 NVIDIA A100-PCIE-40GB GPUs without parallelized LLM inference. The resulting standardized dataset is reused across training and downstream evaluation.

\noindent \textbf{SuPreME Pre-training.}
SuPreME itself consists only of a ViT-based ECG encoder, query-based supervision, and a lightweight Cardiac Fusion Network (CFN). Training is efficient, around 1.5 hours on 4 NVIDIA A100-PCIE-40GB GPUs achieving best AUC performance (16 epochs), and does not require contrastive sampling or further fine-tuning in deployment.

\noindent \textbf{Deployment and Inference}
Once pre-trained, SuPreME supports zero-shot ECG classification via a set of concise cardiac query prompts. Inference only involves a forward pass through the ECG encoder and CFN, taking milliseconds per ECG sample. No LLMs or textual reports are needed at test time, making SuPreME highly practical for deployment in real-world clinical settings. We empirically verify that inference can be efficiently performed on a single NVIDIA A5000-PCIE-24GB GPU or NVIDIA RTX4090-PCIE-24GB GPU.

\subsubsection{Similarity Threshold Determination}
\label{app:dis-threshold}

The similarity thresholds in our entity deduplication and mapping pipeline were determined in consultation with experienced cardiologists (over 10 years of clinical practice), based on joint analysis of the results under various threshold settings in each phase.

Through this process, we observed that setting the thresholds too high (e.g., above 0.9 in entity mapping) would exclude valid clinical variants due to minor wording differences, while setting them too low (e.g., below 0.7 in entity mapping) could introduce semantic ambiguity by incorrectly matching unrelated conditions (Table~\ref{tab:similarity-mismatch}).

\begin{table}[htbp]
    \centering
    \resizebox{\linewidth}{!}{
        \begin{tabular}{l|l|c}
            \toprule
            \textbf{Standard Terminology} & \textbf{Report Entity} & \textbf{Similarity Score} \\
            \midrule
            left ventricle hypertrophy    & Ventricular fibrillation & 0.6400 \\
            non-specific ST changes       & ST elevation            & 0.6343 \\
            inferior myocardial infarction & anterior wall abnormality & 0.5717 \\
            \bottomrule
        \end{tabular}
    }
    \caption{Incorrect matching examples between standard terminology and report entities.}
    \label{tab:similarity-mismatch}
\end{table}

"left ventricle hypertrophy" and "ventricular fibrillation" shows a similarity score of 0.64, but are entirely unrelated - one refers to structural enlargement of the left ventricle, while the other refers to a life-threatening arrhythmia. The selected thresholds reflect a balance between preserving clinically meaningful variants and minimizing noise.

\subsubsection{LLM-based NER Evaluation}
\label{app:dis-llm-ner-eval}

Regarding the performance of NER outputs from LLaMA, we conducted a manual evaluation on 200 randomly sampled clinical reports, annotated by physicians with over ten years of experience. Using a token-level BIO tagging scheme~\citep{ramshaw1999text, moryossef2023linguistically}, we found strong agreement between LLaMA and expert annotations, with an F1 score of 95.71\% and a Cohen’s kappa of 0.93~\citep{cohen1960coefficient, ruan2024defining}. These results confirm the reliability of LLaMA for clinical NER and support the validity of our original findings.

\begin{table}[htbp]
    \centering
    \resizebox{0.45\columnwidth}{!}{
        \begin{tabular}{l|c}
            \toprule
            \textbf{Metrics} & \textbf{Value} \\
            \midrule
            Precision & 96.30\% \\
            Recall    & 95.13\% \\
            F1        & 95.71\% \\
            \midrule
            \textbf{Cohen's kappa} & \textbf{0.93} \\
            \bottomrule
        \end{tabular}
    }
    \caption{Physician-annotated evaluation of LLaMA NER performance.}
    \label{tab:metrics}
\end{table}

\subsubsection{Comparing SuPreME with MERL}
\label{app:dis-merl}

\begin{figure}[htb!]
    \includegraphics[width=\columnwidth]{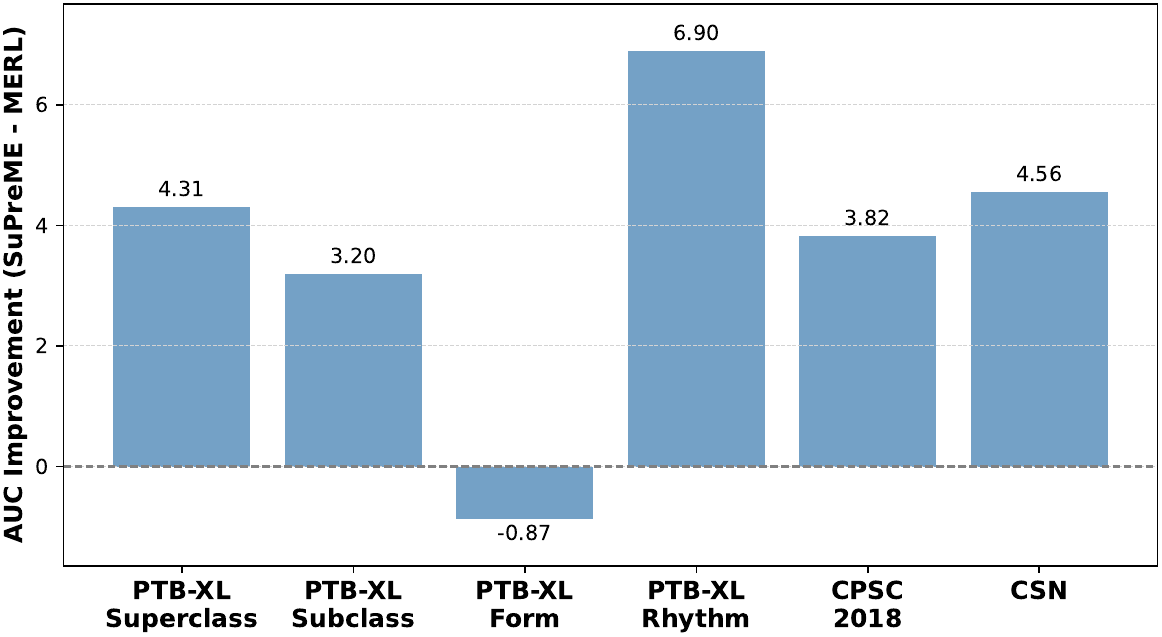}
    \caption{Per-dataset AUC difference between SuPreME and MERL.}
    \label{fig:merl_diff}
\end{figure}

To assess the relative effectiveness of our supervised multimodal framework, we compare SuPreME against MERL~\citep{liu2024zero}, a recent multimodal contrastive learning baseline that utilizes clinical reports and enhanced prompt engineering (Figure~\ref{fig:merl_diff}). While MERL employs contrastive objectives and handcrafted prompts, SuPreME leverages fine-grained diagnostic supervision through LLM-extracted entities and multimodal fusion via the Cardiac Fusion Network (CFN).

\begin{table*}[h]
    \centering
        \resizebox{\linewidth}{!}{
        \begin{tabular}{l|cccccc|c}
            \toprule
            Framework & PTB-XL-Superclass & PTB-XL-Subclass & PTB-XL-Form & PTB-XL-Rhythm & CPSC-2018 & CSN & Avg \\
            \midrule
            MERL & 73.89 & 74.32 & 61.54 & 79.89 & 76.01 & 75.61 & 73.54 \\
            SuPreME & \textbf{78.20} & \textbf{77.52} & 60.67 & \textbf{86.79} & \textbf{79.83} & \textbf{80.17} & \textbf{77.20} \\
            \bottomrule
        \end{tabular}
        }
    \caption{Zero-shot AUC comparison between SuPreME and MERL.}
    \label{tab:merl_comparison}
\end{table*}

As shown in Table~\ref{tab:merl_comparison}, SuPreME achieves consistently higher AUCs across all but one dataset. On average, SuPreME improves zero-shot performance by 3.66\% absolute. Statistical testing confirms this improvement is significant: a paired $t$-test across the six datasets yields $t = 3.51$, $p = 0.0171$.

\begin{table}[h]
    \centering
        \resizebox{0.65\columnwidth}{!}{
            \begin{tabular}{l|c}
                \toprule
                \textbf{Framework} & \textbf{Zero-shot AUC (\%)} \\
                \midrule
                MERL & $73.54 \pm 2.30$ \\
                SuPreME (Ours) & $\mathbf{77.20 \pm 0.21}$ \\
                \bottomrule
            \end{tabular}
        }
    \caption{Average zero-shot AUC and standard deviation across six datasets.}
    \label{tab:zero-shot-std}
\end{table}

Moreover, SuPreME demonstrates significantly lower performance variance across datasets. While MERL exhibits a standard deviation of 2.30, SuPreME achieves a much smaller deviation of 0.21 (Table~\ref{tab:zero-shot-std}), indicating greater robustness and stability across diverse cardiac classification tasks. These results collectively support the effectiveness of our proposed supervised pre-training framework and its entity-level modality fusion strategy, even when compared to a strong multimodal baseline. 

To ensure completeness, we also compare SuPreME with MERL-ResNet in zero-shot settings. Across six datasets, SuPreME achieves higher AUCs on four tasks and yields an overall average gain of +1.95 absolute AUC (equivalent to +2.6\% relative improvement).

\subsubsection{On the Effectiveness of CFN with Different Backbones}
\label{app:dis-cfn-backbone}

To address concerns regarding the effectiveness of the Cardiac Fusion Network (CFN), particularly its relatively lower performance when paired with a ResNet backbone (cf. Table~\ref{tab:zero-shot}), we conduct statistical analyses to better understand the interaction between backbone architecture and the CFN module.

\begin{figure}[htb!]
    \includegraphics[width=\columnwidth]{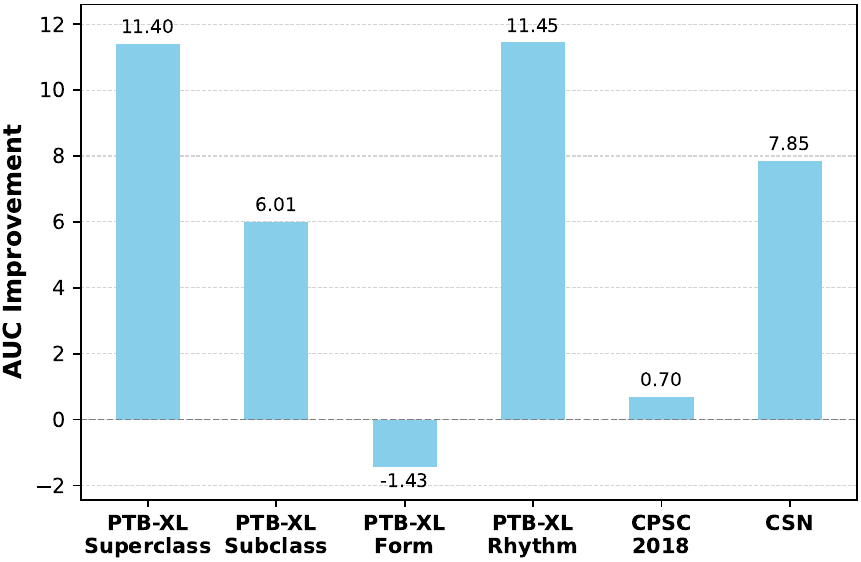}
    \caption{CFN vs. linear classification ($\Delta$ AUC) for ViT ECG backbone.}
    \label{fig:delta_auc}
\end{figure}

\noindent \textbf{$\Delta$ AUC Comparison.}
We compare the AUC improvement brought by CFN over linear classification for both ViT and ResNet backbones across six downstream datasets. As shown in Figure~\ref{fig:delta_auc}, CFN brings consistent performance gains when combined with ViT, with average improvement of +5.97 AUC. In contrast, CFN shows little to negative improvement with ResNet, indicating that the quality of the underlying feature representations plays a critical role in effective cross-modal fusion.

\noindent \textbf{Statistical Significance.}
To verify this trend, we perform a paired t-test and Wilcoxon signed-rank test on the $\Delta$ AUC values. Both tests confirm that CFN yields significantly greater improvements on ViT than ResNet:

\begin{itemize}
    \item \textbf{Paired t-test}: $t = 4.99$, $p = 0.0021$
    \item \textbf{Wilcoxon test}: $W = 21.0$, $p = 0.0156$
\end{itemize}

These results provide strong statistical evidence that ViT synergizes better with CFN compared to ResNet, likely due to ViT’s superior capacity in capturing global temporal dependencies in ECG signals.

CFN is designed to align high-level ECG features with cardiac queries via cross-attention. However, ResNet provides only local, convolutional features with limited contextual depth, especially compared to ViT’s global receptive field. As a result, the decoder lacks sufficient global representations to effectively condition on query semantics. This bottleneck explains the performance drop observed in ResNet + CFN.

\noindent \textbf{Two-Way ANOVA.}
We further conduct a two-way ANOVA with Backbone (ResNet vs. ViT) and Module (Linear vs. CFN) as factors. As shown in Table~\ref{tab:anova}, the interaction term is statistically significant ($F = 6.60$, $p = 0.018$), confirming that the effect of CFN depends on the choice of backbone. Notably, neither factor alone is significant, suggesting that their combination determines performance.

\begin{table}[h]
    \centering
    \resizebox{\columnwidth}{!}{
        \begin{tabular}{lrrrr}
            \toprule
            \textbf{Source} & \textbf{Sum of Squares} & \textbf{df} & \textbf{F-value} & \textbf{p-value} \\
            \midrule
            Backbone                   & 167.06  & 1 & 3.79 & 0.066 \\
            Module                     & 5.57    & 1 & 0.13 & 0.726 \\
            Backbone $\times$ Module   & \textbf{290.65}  & 1 & \textbf{6.60} & \textbf{0.018} \\
            Residual                   & 881.22  & 20 & - & - \\
            \bottomrule
        \end{tabular}
    }
    \caption{Two-way ANOVA results on AUC with backbone and module as factors.}
    \label{tab:anova}
\end{table}

\noindent \textbf{Takeaway.}
These findings reinforce CFN’s role as a powerful fusion mechanism when paired with a backbone (like ViT) that produces expressive feature sequences. The drop in performance with ResNet may stem from its less structured output, which lacks the sequential token-style organization needed for effective query-based attention. Thus, the CFN is not inherently ineffective, but its utility hinges on a compatible encoder design.

\subsubsection{Domain Scope and Generalization Potential}
\label{app:dis-domain}

While our framework is evaluated on 12-lead ECG data, we believe that this modality represents a highly impactful and widely applicable domain in clinical practice. ECG is routinely used across diverse medical contexts, including emergency rooms, intensive care units (ICUs), outpatient cardiology clinics, and even home-based healthcare monitoring, due to its low cost, non-invasiveness, and real-time ability to reflect cardiac electrical activity. As such, improving automated ECG interpretation has direct clinical relevance across resource settings and specialties.

Moreover, although this work focuses on ECG, the core methodology of SuPreME, namely multimodal learning between biomedical signals and clinically meaningful queries, can be generalized to other physiological signal domains such as EEG (electroencephalogram) or PPG (photoplethysmography). These modalities are similarly structured (multi-channel, time-series signals) and increasingly available in clinical and wearable settings. However, to the best of our knowledge, there is currently a lack of large-scale, publicly accessible datasets that pair these signals with detailed, free-text clinical reports suitable for training our entity extraction module.

We hope our work can inspire future efforts toward building such paired datasets for other biomedical signals, enabling the broader application of query-based multimodal learning frameworks beyond ECG.

\end{document}